\documentclass[twocolumn,preprintnumbers,amsmath,amssymb,amsfonts]{revtex4-1}
\usepackage{epsfig}
\usepackage{graphicx}

\begin{document}
\preprint{UTTG-15-10}
\preprint{TCC-034-10}

\title{Non-Gaussianities in Multifield Inflation: Superhorizon Evolution, Adiabaticity, and the Fate of $f_{\mathrm{NL}}$}

\author{Joel Meyers}
\email{joelmeyers@mail.utexas.edu}
\author{Navin Sivanandam}
\email{navin.sivanandam@gmail.com}
\affiliation{Theory Group, Department of Physics, University of Texas, Austin, TX 78712}
\affiliation{Texas Cosmology Center, University of Texas, Austin, TX 78712}

\date{\today}

\begin{abstract}
We explore the superhorizon generation of large $f_{\mathrm{NL}}$ of the local form in two field inflation. We calculate the two- and three-point observables in a general class of potentials which allow for an analytic treatment using the $\delta N$ formalism. Motivated by the conservation of the curvature perturbation outside the horizon in the adiabatic mode and also by the observed adiabaticity of the power spectrum, we follow the evolution of $f_{\mathrm{NL}}^{\textrm{local}}$ until it is driven into the adibatic solution by passing through a phase of effectively single field inflation.  We find that although large $f_{\mathrm{NL}}^{\textrm{local}}$ may be generated during inflation, such non-gaussianities are transitory and will be exponentially damped as the cosmological fluctuations approach adiabaticity.
\end{abstract}

\maketitle

\section{Introduction}
Amongst the striking successes of cosmic inflation, the most dramatic is the natural prediction of the cosmological fluctuations observed in the cosmic microwave background and in large scale structure \cite{Mukhanov:1981xt,Hawking:1982cz,Starobinsky:1982ee,Guth:1982ec,Bardeen:1983qw,Fischler:1985ky}.  Many models of inflation predict a nearly scale-invariant, nearly gaussian power spectrum which accords with current observation, though measurements of the scalar spectral index and bounds on the relative gravitational wave amplitude have begun to rule out some models of inflation \cite{Komatsu:2010fb}.  While ongoing and future satellite experiments will greatly improve the precision of measurements of the power spectrum \cite{Perotto:2006rj,Hamann:2007sb} (and thus help to further constrain theories of the early universe), there is a limit to the insight gained from the power spectrum alone \cite{Easson:2010zy,Easson:2010uw}.

To go further, we can consider the non-gaussian parts of the spectrum, which in principle contain a much richer set of information, and thus may provide a key observational tool to distinguish between detailed models of inflation. In particular, there has been a great deal of progress in measuring the bispectrum (3-point function) from the CMB \cite{Komatsu:2010fb} and from large scale structure \cite{Slosar:2008hx}. Complete characterization of the 3-point function is a difficult task, but a useful basis can be given in terms of different shapes of momentum-space triangles. In particular, one considers the following forms (we define $k_1\geq k_2\geq k_3$ in the following): local/squeezed ($k_1\sim k_2\gg k_3$), equilateral ($k_1\sim k_2\sim k_3$) and orthogonal (constructed to be nearly orthogonal to the preceding two forms) -- see \cite{Komatsu:2010hc} for a review of the theoretical and observational motivation for this basis. Of course, these are not the only forms one can write down, and other shapes may probe other aspects of inflation.

The current observational bounds are $-5<f_{\mathrm{NL}}^{\textrm{local}}<59$ (WMAP7+SDSS), $-214<f_{\mathrm{NL}}^{\textrm{equil}}<266$ (WMAP7) and $-410<f_{\mathrm{NL}}^{\textrm{orthog}}<6$ (WMAP7), where $f_{\mathrm{NL}}$ characterizes the size of the bispectrum compared with the power spectrum (see (\ref{fnldef}) for the precise definition).

The local shape is especially important because it is predicted to be small in all models of single field inflation \cite{Maldacena:2002vr,Creminelli:2004yq,Ganc:2010ff,RenauxPetel:2010ty}, so a convincing detection of significant local-shape bispectrum would therefore rule out a wide class of models including the simplest single field inflationary scenario. This result does not, however, hold for multifield inflation; in particular there are a number of models that use the continued evolution of the curvature perturbation outside the horizon in order to generate a large $f_{\mathrm{NL}}^{\textrm{local}}$; for example see \cite{Linde:1996gt,Bernardeau:2002jy,Bernardeau:2002jf,Lyth:2002my,Wands:2010af,Byrnes:2009qy,Battefeld:2006sz,Chen:2009we,Sasaki:2008uc,Gao:2008dt,Battefeld:2009ym,Byrnes:2008zy,Cai:2009hw,Rigopoulos:2005us,Rigopoulos:2005ae}.

This superhorizon evolution of the curvature perturbation is characteristic of the presence of non-adiabatic fluctuations in the cosmological fluid, which is itself one of the most important differences between single field inflation and models with multiple dynamical fields. Since, regardless of the contents of the universe, there always exist two purely adiabatic modes in which the superhorizon curvature perturbation is conserved \cite{Weinberg:2003sw, Weinberg:2008nf}, the (only) two scalar fluctuation modes in single field inflation are guaranteed to be adiabatic and remain so for the subsequent evolution of the universe \cite{Weinberg:2004kr}. This means that for a given model of single field inflation, we need only calculate the evolution of the perturbations until the time of horizon exit in order to make predictions about the observational effects of cosmological fluctuations.  Models with multiple dynamical fields, however, admit non-adiabatic solutions in which the curvature perturbation can evolve outside the horizon.  As a result, we must calculate the evolution of the curvature perturbation until it becomes conserved, or until it is observed.  If the curvature perturbation does not become conserved, models with multiple dynamical fields cannot make sharp predictions about cosmological fluctuations unless we have a complete understanding of the entire history of the universe stretching back to the time of inflation.

If there are non-adiabatic fluctuations present in the cosmological fluid during the radiation-dominated era this would leave observable effects on the cosmic microwave background.  Specifically, these fluctuations carry a different phase than adiabatic fluctuations, and have peaks at different positions in the angular power spectrum \cite{Bucher:1999re,Bucher:2000kb,Bucher:2000cd,Bucher:2000hy}.  Measurements of the power spectrum can place upper limits on the magnitude of the non-adiabatic contribution to the spectrum of fluctuations in the early universe, and current observations have not found any evidence for such a contribution \cite{Komatsu:2010fb}.

The non-observation of a non-adiabatic (or isocurvature) contribution to the spectrum is not by itself enough to rule out multifield inflation, since there are at least two ways in which the perturbations produced by multifield inflation can become adiabatic. The first is to pass through a phase of effectively single field inflation, by which we mean that all but one of the eigenvalues of the mass matrix for the inflaton fields are large and positive.  Secondly, the adiabatic solution is attractive if the universe passes through a sufficiently long period of local thermal equilibrium with no non-zero conserved quantum numbers \cite{Weinberg:2004kf,Weinberg:2008si}.

Now, as observed above, the principal suggestions for having a large $f_{\mathrm{NL}}^{\textrm{local}}$ generated in a multifield model use the evolution of the curvature perturbation outside the horizon in the presence of non-adiabatic fluctuations to boost a initially small bispectrum. In this way, superhorizon evolution of the curvature perturbation allows for the generation of non-gaussianity, even if fluctuations were nearly gaussian at horizon exit.

While the above scenario is apparently promising with regards to the generation of large $f_{\mathrm{NL}}^{\textrm{local}}$, it is also incomplete. If there remain non-adiabatic fluctuations in the cosmological fluid, the curvature perturbation and its correlations (including the bispectrum) continue to evolve. Thus any calculated non-gaussianity in a particular multifield model is completely unpredictive unless it can be carried through to an epoch in which the perturbations either become purely adiabatic (i.e. a phase of effectively single field inflation or a period of local thermal equilibrium) or are observed.

Our goal in this paper, then, is to study the continued evolution of $f_{\mathrm{NL}}^{\textrm{local}}$ until the curvature perturbations becomes constant and thus address whether there exist models of multiple field inflation which predict observably large local non-gaussianity and a purely adiabatic power spectrum. We insist on the latter not only to be consistent with observational bounds, but also to be certain that any alleged non-gaussianity would survive to the present epoch.

To this end we use the $\delta N$ formalism to obtain analytical formulas for two- and three-point statistics in a wide class of potentials for two-field inflation.  We then require that the fluctuations become purely adiabatic before inflation ends by passing through a phase of effectively single-field inflation.  For the potentials we study, the suppression of non-adiabatic fluctuations by a phase of single field inflation also dynamically suppresses the local bispectrum.  We therefore conclude that for this set of models, regardless of initial conditions, a purely adiabatic power spectrum and large local non-gaussianity cannot be simultaneously produced from inflation alone.

Our conclusions about $f_{\mathrm{NL}}^{\textrm{local}}$ seem to be in agreement with numerical studies of particular models of multifield inflation; see for example \cite{Vernizzi:2006ve,Battefeld:2009ym,Wanatanabe:2009}.

Of course, the space of models that could produce non-gaussianities and an adiabatic power spectrum is considerably larger than the set we examine. In particular there are models such as the curvaton scenario and modulated reheating, where the non-adiabatic fluctuations persist through the end of inflation and are damped away during a period of local thermal equilibrium. It is reasonable to suggest that in these models, or in a different class of potentials, non-gaussianities generated from superhorizon evolution of the curvature perturbation may be observable. We will discuss these possibilities below and will be exploring them in more detail in subsequent work. However, we believe that without understanding the complete evolution of perturbations until they become adiabatic, it is somewhat premature to claim observable $f_{\mathrm{NL}}^{\textrm{local}}$ as a prediction of multifield inflation.

The structure of the paper is as follows. In section \ref{themodel} we review the $\delta N$ formalism, find the form of the most general two dimensional potential to which such formalism can be readily applied and derive expressions for the statistical properties of the curvature fluctuations for two particular forms of our potential (the painful details of the general case are relegated to appendix \ref{limitedtraj}). In section \ref{fnlfate}, for the same two potential forms (again the general case can be found in appendix \ref{limitedtraj}), we show that a phase of single field inflation drives $f_{\mathrm{NL}}^{\textrm{local}}$ to be slow roll suppressed. In section \ref{evading} we discuss how our results may be circumvented. And, finally, we summarize our conclusions in section \ref{conclusion}.

\section{The Model}\label{themodel}
\subsection{$\delta N$ and the Spectrum}
Consider an action of the form:
\begin{equation}
S = \int\textrm{d}^4x\sqrt{-g}\left[\frac{1}{2}m_p^2R+\frac{1}{2}g^{\mu\nu}G_{ab}\partial_{\mu}\phi^a\partial_{\nu}\phi^b-W(\vec{\phi})\right] \, .
\end{equation}
We concern ourselves with two field models with canonical kinetic terms, though, as we note below, some extensions to more fields will not substantially change our central conclusions. In this case $G_{ab}=\delta_{ab}$, with $a,b=1,2$, and for clarity we use $\phi$ and $\chi$ for the two scalar fields. The equations of motion are then given by:
\begin{align}\label{eqm}
0&=\ddot{\phi}+3H\dot{\phi}+\partial_{\phi}W\nonumber\\
0&=\ddot{\chi}+3H\dot{\chi}+\partial_{\chi}W\nonumber\\
H^2&=\frac{1}{3m_p^2}\left(\frac{1}{2}\dot{\phi}^2+\frac{1}{2}\dot{\chi}^2+W\left(\phi,\chi\right)\right)\nonumber\\
\dot{H}&=-\frac{1}{2m_p^2}\left(\dot{\phi}^2+\dot{\chi}^2\right)\, .
\end{align}
Assuming slow roll ($|\dot{H}|\ll H^2$ and $|\ddot{\phi_a}|\ll H|\dot{\phi_a}|$) these become:
\begin{align}\label{eqm1}
3H\dot{\phi}&\simeq-\partial_{\phi}W\nonumber\\
3H\dot{\chi}&\simeq-\partial_{\chi}W\nonumber\\
H^2&\simeq\frac{1}{3m_p^2}W\, .
\end{align}

Models of this type and their associated non-gaussianities have been well studied in the literature, for specific classes of potentials -- for example sum-separable potentials (e.g. \cite{Vernizzi:2006ve}), product-separable potentials (e.g. \cite{GarciaBellido:1995qq,Choi:2007su}) and many others (e.g. \cite{Wands:2010af,Byrnes:2009qy,Battefeld:2006sz,Chen:2009we,Sasaki:2008uc,Gao:2008dt,Battefeld:2009ym,Byrnes:2008zy}). Such studies mostly use the $\delta N$ formalism \cite{Starobinsky:1986fxa,Sasaki:1995aw,Lyth:2004gb,Lyth:2005fi} to calculate the evolution of the curvature perturbation, $\zeta$. Accordingly, we begin by reviewing this formalism -- our treatment follows closely that of Vernizzi and Wands \cite{Vernizzi:2006ve}.

$\zeta$ is defined as the curvature perturbation on uniform density hypersurfaces \cite{Bardeen:1980kt,Bardeen:1983qw} and is of particular utility, since it is conserved for adiabatic modes (i.e. those from single field inflation) whose physical size is greater than that of the horizon. In terms of the perturbed metric (tensor and vector perturbations are suppressed):
\begin{multline}
ds^2=-(1+2A)dt^2+2B_{,i}dtdx\\
+a(t)^2\left[\left(1-2\psi\right)\delta_{ij}+2E_{,ij}\right]dx^idx^j\, ,
\end{multline}
$\zeta$ is given up to second order by:
\begin{equation}
\zeta\equiv-\psi-\psi^2-\frac{H}{\dot{\rho}}\delta\rho+\frac{1}{\dot{\rho}}\dot{\psi}\delta\rho+\frac{H}{\dot{\rho}^2}\delta\rho\dot{\delta\rho}+\frac{1}{2\dot{\rho}}\left(\frac{H}{\dot{\rho}}\right)^.\delta\rho^2\, ,
\end{equation}
where $\rho$ is the energy density and $\delta\rho$ is its perturbation.

The heart of the $\delta N$ formalism is the observation that, at large scales, $\zeta$ (at some comoving time $t_c$) is given by the perturbation to the number of e-foldings from an initially flat hypersurface (at $t=t_*$) to a comoving one (at $t=t_c$):
\begin{equation}\label{Nzeta}
\zeta\left(t_c,\vec{x}\right)\simeq\delta N\left(t_c,t_*,\vec x\right)\equiv\mathcal{N}\left(t_c,t_*,\vec x\right)-N\left(t_c,t_*\right)\, .
\end{equation}
$N$ is the unperturbed number of e-foldings, given by integrating $H$ from $t_*$ to $t_c$:
\begin{equation}\label{Nint}
N=\int_*^cH\textrm{d}t \, .
\end{equation}

If we take $t_*$ as the time a particular mode exits the horizon (i.e. when $k=aH$), the number of e-foldings can be viewed as a function of the field configuration on the hypersurface defined by horizon exit, $\phi^I(t_*,\vec{x})$, and $t_c$. The perturbation in $N$ can then be expressed (up to some order, in our case second) in terms of the fluctuations of the scalar fields at horizon exit:
\begin{equation}\label{deltaN}
\delta N\simeq\sum_I N_{,I}\delta\phi_*^I+\sum_{IJ}N_{,IJ}\delta\phi_*^I\delta\phi_*^J\, .
\end{equation}
The derivatives here are with respect to the fields at $t=t_*$ ($N,_I \equiv \frac{\partial N}{\partial \phi_*^I}$). In reality, $\delta N$ should also depend on the values of the scalar field velocities. However, so long as the slow roll approximation holds at horizon exit, only the field values determine the subsequent dynamics.

\subsubsection*{Statistics}
The utility of the $\delta N$ formalism is that it provides a readily tractable way to calculate cosmological observables; in particular, two and three-point statistics. We summarize the relevant results below, again we closely follow the treatment of Vernizzi and Wands \cite{Vernizzi:2006ve} (see also \cite{Seery:2005gb}).

The power spectrum, $\mathcal{P}_{\zeta}$ is defined by:
\begin{equation}\label{Pzeta}
\left\langle\zeta_{\boldsymbol{k}_1}\zeta_{\boldsymbol{k}_2}\right\rangle\equiv(2\pi)^3\delta^{(3)}(\boldsymbol{k}_1+\boldsymbol{k}_2)\frac{2\pi^2}{k_1^3}\mathcal{P}_{\zeta}(k_1)\, .
\end{equation}
In a similar fashion the two-point correlation function for the scalar field fluctuations at horizon exit is given by:
\begin{align}\label{Pstar}
\left\langle\delta\phi^I_{\boldsymbol{k}_1}\delta\phi^J_{\boldsymbol{k}_2}\right\rangle&\equiv(2\pi)^3\delta^{IJ}\delta^{(3)}(\boldsymbol{k}_1+\boldsymbol{k}_2)\frac{2\pi^2}{k_1^3}\mathcal{P}_*(k_1)\, , \nonumber\\
P_*(k)&\equiv\frac{H_*^2}{4\pi^2}\, .
\end{align}
Then, from (\ref{Nzeta}), (\ref{deltaN}), (\ref{Pzeta}) and (\ref{Pstar}), we have:
\begin{equation}\label{PN}
\mathcal{P}_{\zeta}=\sum_IN_{,I}^2\mathcal{P}_*\, .
\end{equation}
The spectral index is given by (the approximate equality denotes lowest order in slow roll):
\begin{equation}
n_{\zeta}-1\equiv\frac{\textrm{d}\ln\mathcal{P}_{\zeta}}{\textrm{d}\ln k}\simeq\frac{\textrm{d}\ln\mathcal{P}_{\zeta}}{\textrm{d}N}=\frac{1}{H}\frac{\textrm{d}\ln\mathcal{P}_{\zeta}}{\textrm{d}t}\, .
\end{equation}
Using the above definition and (\ref{PN}) we have:
\begin{equation}\label{ns1}
n_{\zeta}-1=-2\epsilon+\frac{2}{H}\frac{\sum_{IJ}\dot{\phi}_JN_{,IJ}N_{,I}}{\sum_KN^2_{,K}}\, .
\end{equation}
$\epsilon\equiv-\dot{H}/H^2$ is the usual slow roll parameter. Using the slow roll equations of motion, this can be written as \cite{Lyth:1998xn}:
\begin{equation}\label{ns2}
n_{\zeta}-1=-2\epsilon-\frac{2}{m_p^2\sum_KN^2_{,K}}+\frac{2m_p^2\sum_{IJ}W_{,IJ}N_{,I}N_{,J}}{W\sum_KN^2_{,K}}\, .
\end{equation}

Three point statistics can be obtained in much the same fashion. The curvature bispectrum, $B_{\zeta}$, is defined through:
\begin{equation}\label{Bzeta}
\left\langle\zeta_{\boldsymbol{k}_1}\zeta_{\boldsymbol{k}_2}\zeta_{\boldsymbol{k}_3}\right\rangle\equiv(2\pi)^3\delta^{(3)}\left(\sum_i\boldsymbol{k}_i\right)\mathcal{B}_{\zeta}(k_1,k_2,k_3)\, .
\end{equation}
The bispectrum can be used to define the non-linearity parameter $f_{\mathrm{NL}}$ (defined in \cite{Maldacena:2002vr}) which is the quantity most often referenced in observational constraints:
\begin{equation}\label{fnldef}
\frac{6}{5}f_{\mathrm{NL}}\equiv\frac{\prod_ik_i^3}{\sum_ik_i^3}\frac{B_{\zeta}}{4\pi^4\mathcal{P}_{\zeta}^2}\, .
\end{equation}
From (\ref{Nzeta}) and (\ref{deltaN}):
\begin{multline}\label{3ptN}
\left\langle\zeta_{\boldsymbol{k}_1}\zeta_{\boldsymbol{k}_2}\zeta_{\boldsymbol{k}_3}\right\rangle=\sum_{IJK}N_{,I}N_{,J}N_{,K}\left\langle\delta\phi^I_{\boldsymbol{k}_1}\delta\phi^J_{\boldsymbol{k}_2}\delta\phi^K_{\boldsymbol{k}_3}\right\rangle\\
+\frac{1}{2}\sum_{IJKL}N_{,I}N_{,J}N_{,KL}\left\langle\delta\phi^I_{\boldsymbol{k}_1}\delta\phi^J_{\boldsymbol{k}_2}(\delta\phi^K\star\delta\phi^L)_{\boldsymbol{k}_3}\right\rangle\\
+\textrm{perms}\, .
\end{multline}
The star denotes a convolution and higher order terms have been neglected \cite{Zaballa:2006pv,Seery:2005gb}.

The three-point field correlation for the fields is given by \cite{Seery:2005gb}:
\begin{multline}
\left\langle\delta\phi^I_{\boldsymbol{k}_1}\delta\phi^J_{\boldsymbol{k}_2}\delta\phi^K_{\boldsymbol{k}_3}\right\rangle=\\
(2\pi)^3\delta^{(3)}\left(\sum_i\boldsymbol{k}_i\right)\frac{4\pi^4}{\prod_ik_i^3}\mathcal{P}_*^2\sum_{\textrm{perms}}\frac{\dot{\phi}_I\delta_{IJ}}{4Hm_p^2}\mathcal{M}\, .
\end{multline}
The sum is over simultaneous rearrangements of $I$, $J$ and $K$ and the momenta $k_1$, $k_2$ and $k_3$ and $\mathcal{M}$ is given by:
\begin{align}
\mathcal{M}(k_1,k_2,k_3)&\equiv-k_1k_2^2-4\frac{k_2^2k_3^2}{k_t}+\frac{1}{2}k_1^3+\frac{k_2^2k_3^2}{k_t^2}(k_2-k_3)\, , \nonumber\\
k_t&=k_1+k_2+k_3\, .
\end{align}
Carrying our the sum over permutations gives the first term in (\ref{3ptN}) \cite{Vernizzi:2006ve} as (using $\sum_IN_{,I}\dot{\phi}_I=H$):
\begin{multline}
\sum_{IJK}N_{,I}N_{,J}N_{,K}\sum_{\textrm{perms}}\dot{\phi}_I\delta_{JK}\mathcal{M}(k_1,k_2,k_3)=\\
-H\sum_IN_{,I}^2\mathcal{F}(k_1,k_2,k_3)\, ,
\end{multline}
where:
\begin{align}
\mathcal{F}(k_1,k_2,k_3)&\equiv\sum_{\textrm{perms}}\mathcal{M}(k_1,k_2,k_3)\nonumber\\
&=-2\left(\frac{1}{2}\sum_{i\neq j}k_ik_j^2+4\frac{\sum_{i>j}k_i^2k_j^2}{k_t}-\frac{1}{2}\sum_ik_i^3\right)\, .
\end{align}

For the second term in (\ref{3ptN}), we follow \cite{Vernizzi:2006ve,Seery:2005gb} and neglect the connected part of the four-point function and use Wick's theorem to obtain (after a few manipulations):
\begin{multline}
B(k_1,k_2,k_3)=\\
4\pi^4\mathcal{P}_{\zeta}^2\frac{\sum_ik_i^3}{\prod_ik_i^3}\left(\frac{-1}{4m_p^2\sum_KN_{,K}^2}\frac{\mathcal{F}}{\sum_ik_i^3}\right.\\
\left.+\frac{\sum_{IJ}N_{,I}N_{,J}N_{,IJ}}{\left(\sum_KN_{,K}^2\right)^2}\right)\, .
\end{multline}
From the above and (\ref{fnldef}), $f_{\mathrm{NL}}$ is:
\begin{equation}\label{fnltotal}
\frac{6}{5}f_{\mathrm{NL}}=\frac{\mathcal{P}_*}{2m_p^2\mathcal{P}_{\zeta}}(1+f)+\frac{\sum_{IJ}N_{,I}N_{,J}N_{,IJ}}{(\sum_KN_{,K}^2)^2}\, .
\end{equation}
The function $f$ is given by:
\begin{equation}
f(k_1,k_2,k_3)\equiv-1-\frac{\mathcal{F}}{2\sum_ik_i^3}\, ,
\end{equation}
and is a shape-dependent (in momentum space) function, with values between 0 and 5/6 \cite{Maldacena:2002vr}. Then, defining  the scalar-tensor ratio $r\equiv8\mathcal{P}_*/m_p^2\mathcal{P}_{\zeta}$, we can rewrite the first term of (\ref{fnltotal}) as:
\begin{equation}
\frac{6}{5}f_{\mathrm{NL}}^{(3)}\equiv\frac{r}{16}(1+f)\, ,
\end{equation}
which constrained by observations to be small since $r\ll1$ \cite{Komatsu:2010fb}.

The second term in (\ref{fnltotal}), $f_{\mathrm{NL}}^{(4)}$, is momentum-independent and local in real space (so it contributes to $f_{\mathrm{NL}}^{\mathrm{local}}$), and can (in principle at least) be larger than unity:
\begin{equation}
\frac{6}{5}f_{\mathrm{NL}}^{(4)}\equiv\frac{\sum_{IJ}N_{,I}N_{,J}N_{,IJ}}{(\sum_KN_{,K}^2)^2}\, .
\end{equation}
The current bounds are $-5<f_{\mathrm{NL}}^{\textrm{local}}<59$ (WMAP7+SDSS), and these should tighten considerably with the current generation of CMB experiments, with the Planck satellite expected to give $\Delta f_{\mathrm{NL}}^{\textrm{local}}\sim5$ \cite{Komatsu:2001rj}. As noted above $f_{\mathrm{NL}}^{\textrm{local}}$ is slow roll suppressed in all models of single field inflation and thus could be (in principle at least) a sharp probe of the number of dynamical scalar degrees of freedom during inflation. However, as we shall demonstrate below, generating and preserving a large $f_{\mathrm{NL}}^{\textrm{local}}$ through to adiabaticity is far from easy.

\subsection{The Potential}\label{potentialdetails}
In order for the above expressions to be useful we need to take derivatives of $N$ with respect to the initial values of the fields. In multifield inflation this is not straightforward, for, unlike in single field inflation, there are an infinite number of possible inflationary trajectories. As a result, changes in the initial conditions affect not only how far along an inflationary trajectory the field has traveled at some later time, but also the particular trajectory the field is actually on. This easiest way to deal with this difficulty is to find a constant of the motion that allows us to relate the initial and final field values to one another -- in much the same way as conservation of energy and momentum in particle mechanics allow us to connect initial and final states without having to know the detailed form of a particular solution.

\begin{figure}[t]
\centering
\includegraphics[width=\columnwidth]{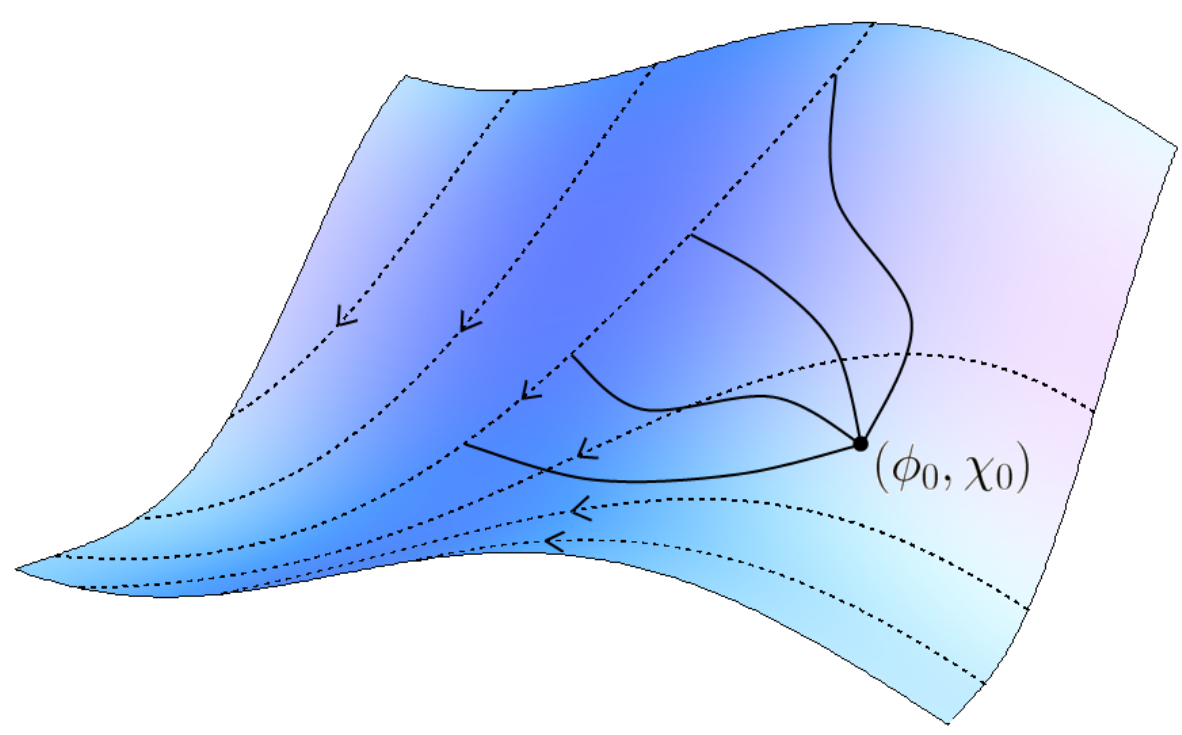}
\caption{The dashed lines show possible classical paths of the inflaton ($N$ increases along the path). The quantity $C$, defined as an intergral along the solid lines emanating from the point $(\phi_0,\chi_0)$, depends only on the endpoints of the curve and does not change as one moves along a trajectory.}
\label{Ctraj}
\end{figure}

To construct such a constant we define a vector $\vec{A}$ (in field space) and consider the quantity:
\begin{equation}
C_{\gamma}=\int_{\gamma}\vec{A}\cdot\textrm{d}\vec{\phi}\, .
\end{equation}
The integral is defined along a path $\gamma$ in field space. In order for $C_{\gamma}$ to be a good constant of motion it needs to be independent of the particular path $\gamma$ (and of course constant along trajectories); figure \ref{Ctraj} illustrates the construction of $C$. In this case we have:
\begin{align}
C_{\gamma}&=\int_{\gamma}\vec{A}\cdot\textrm{d}\vec{\phi}\nonumber\\
&=\int_{\gamma}\vec{\nabla}\Psi\cdot\textrm{d}\vec{\phi}\, .
\end{align}
The derivative of $C$ is then given by:
\begin{equation}
\frac{\textrm{d}C}{\textrm{d}t}=\partial_{\phi}\Psi\dot{\phi}+\partial_{\chi}\Psi\dot{\chi}
\end{equation}
Setting this equal to 0 and using the slow roll equations of motion (\ref{eqm1}) gives:
\begin{equation}
\partial_{\phi}\Psi\partial_{\phi}W=-\partial_{\chi}\Psi\partial_{\chi}W\, .
\end{equation}
and thus we can write:
\begin{align}\label{WPsi}
\partial_{\phi}\Psi&=\frac{g(\phi,\chi)}{\partial_{\phi}W} & \partial_{\chi}\Psi&=-\frac{g(\phi,\chi)}{\partial_{\chi}W}\, .
\end{align}
For such a $\Psi$ to exist, we must have (subscripts denote derivatives):
\begin{equation}\label{Wg}
\frac{g_{\chi}}{W_{\phi}}-\frac{gW_{\phi\chi}}{W_{\phi}^2}=-\frac{g_{\phi}}{W_{\chi}}+\frac{gW_{\phi\chi}}{W_{\chi}^2}\, .
\end{equation}
In principle any $W$ satisfying the above equation for an arbitrary $g$ would allow us to define a suitable $C$. However, if we wish derivatives of $C$ to only depend on the choice of trajectory and not the fiducial point at which our integral is anchored ($(\phi_0, \chi_0)$ in figure \ref{Ctraj}) then $\partial_{\phi}\Psi$ needs to be only a function of $\phi$ and likewise $\partial_{\chi}\Psi$ needs to be only a function of $\chi$ so that the dependence on endpoints in $C$ is separable: $C=C_1(\phi(t),\chi(t))+C_2(\phi_0,\chi_0)$. This is turn means both the left and right-hand sides of (\ref{Wg}) are identically zero.

We can solve the above equations with $W(\phi,\chi)\equiv F(U(\phi)+V(\chi))$ and $g=F'$. We stress that this is not necessarily the only potential for which a suitable constant of motion can be found, but it is the most general one we have been able to construct. With $W$ of this form, $C$ is given by:
\begin{equation}\label{Cpotential}
C=-m_p^2\int_{\phi_0}^{\phi}\frac{1}{U'(\phi')}\textrm{d}\phi'+m_p^2\int_{\chi_0}^{\chi}\frac{1}{V'(\chi')}\textrm{d}\chi'\, .
\end{equation}

We note at this juncture that this class of potentials has also been recently analyzed in \cite{Wang:2010si} by Wang (who also considers  a possible non-canonical kinetic term for the $\chi$ field). Our analysis differs somewhat in the particular approach to the calculation.

With the above form of $W$ the slow roll equations of motions can be written as:
\begin{align}
3H\dot{\phi}&\simeq F'U'\nonumber\\
3H\dot{\chi}&\simeq F'V'\nonumber\\
H^2&\simeq\frac{1}{3m_p^2}W\left(\phi,\chi\right)\, .
\end{align}
Here, as below, a prime denotes the derivative with respect the argument of the function. The slow roll parameters ($\epsilon_i=(m_p^2/2)(\partial W/\partial\phi_i)^2$ and $\eta_{ij}=m_p^2(\partial^2W/\partial\phi_i\phi_j)$) are given by (we drop the repeated index on the diagonal $\eta$ for conciseness):
\begin{align}\label{slowroll}
\epsilon^{\phi}&\equiv\frac{m_p^2}{2}\left(\frac{F'U'}{F}\right)^2\nonumber\\
\epsilon^{\chi}&\equiv\frac{m_p^2}{2}\left(\frac{F'V'}{F}\right)^2\nonumber\\
\epsilon&=\epsilon^{\phi}+\epsilon^{\chi}=-\frac{\dot{H}}{H^2}\nonumber\\
\eta^{\phi}&\equiv m_p^2\left(\frac{F''U'^2+F'U''}{F}\right)\nonumber\\
\eta^{\chi}&\equiv m_p^2\left(\frac{F''V'^2+F'V''}{F}\right)\nonumber\\
\eta^{\phi\chi}&\equiv m_p^2\left(\frac{F''U'V'}{F}\right)=2\frac{FF''}{F'^2}\sqrt{\epsilon^{\phi}\epsilon^{\chi}}\, .
\end{align}

As noted above, $N$ is given by the integral:
\begin{equation}
N=\int_*^cH\rm{d}t\, .
\end{equation}
In order to be able to take derivatives of $N$ with respect to $\phi_*$ and $\chi_*$ (necessary to use the $\delta N$ formalism) we use the slow roll equations of motion to rewrite the above expression for $N$ as:
\begin{multline}\label{Npotential}
N=-\frac{1}{2m_p^2}\int_{*}^{c}\frac{W\left(\phi,\chi(\phi)\right)}{W_{\phi}\left(\phi,\chi(\phi)\right)}\textrm{d}\phi\\
-\frac{1}{2m_p^2}\int_{*}^{c}\frac{W\left(\chi,\phi(\chi)\right)}{W_{\chi}\left(\chi,\phi(\chi)\right)}\textrm{d}\chi\, .
\end{multline}
We've explicitly shown the dependence of each of the fields on the other to emphasize that the integrals can only be performed for a subset of all possible trajectories -- those for which there is a one-to-one mapping from $\phi$ to $\chi$. Even with this restriction, the calculation of the derivatives of $N$ is somewhat involved; for completeness we include the gory details in appendix \ref{limitedtraj} but limit ourselves to a simpler (more restricted) class of models in the main body of this work.

The aforementioned simplification comes from further constraining the form of the potential, so that $N$ separates into distinct integrals over $\phi$ and $\chi$. To do this we rewrite (\ref{Npotential}) as:
\begin{equation}
N=-\frac{1}{m_p^2}\int_{*}^{c}\frac{1}{U'(\phi)}\frac{F\left(U(\phi)+V(\chi)\right)}{F'\left(U(\phi)+V(\chi)\right)}\textrm{d}\phi\, .
\end{equation}
Then, if we require that $F$ is of a form such that:
\begin{equation}
\frac{F\left(U(\phi)+V(\chi)\right)}{F'\left(U(\phi)+V(\chi)\right)}\equiv \tilde{U}(\phi)+\tilde{V}(\chi)\, ,
\end{equation}
a little manipulation gives:
\begin{equation}
N=-\frac{1}{m_p^2}\int_{*}^{c}\frac{1}{U'(\phi)}\tilde{U}(\phi)\textrm{d}\phi-\frac{1}{m_p^2}\int_{*}^{c}\frac{1}{V'(\chi)}\tilde{V}(\chi)\textrm{d}\chi
\end{equation}

One can show that the most general solution for $F$ of this form is given by \footnote{If $f(a+b)\equiv g(a)+h(b)$, then $g(a)-h(a)$ is a constant which in turn means $f(a+b)\equiv g(a)+g(b)+c$ so that $g$ and $f$ are identical up to constant, and thus $f$ is a linear function.}:
\begin{equation}
\frac{F\left(U(\phi)+V(\chi)\right)}{F'\left(U(\phi)+V(\chi)\right)}=\alpha\left(U(\phi)+V(\chi)\right)+2\beta\, ,
\end{equation}
with $\alpha$, $\beta$ constants. With $F$ constrained thusly, the most general form of $W$ is given by:
\begin{equation}
W=W_0\left[\alpha\left(U(\phi)+V(\chi)\right)+2\beta\right]^{1/\alpha}\, .
\end{equation}
With suitable redefinitions of parameters and potentials this can be rewritten as ($\alpha\neq0$):
\begin{equation}
W=\left[U(\phi)+V(\chi)\right]^{\gamma}\, .
\end{equation}
So that $N$ is then given by:
\begin{equation}\label{Nhomogeneous}
N=-\frac{1}{\gamma m_p^2}\int_*^c\frac{U(\phi)}{U'\left(\phi\right)}\textrm{d}\phi
-\frac{1}{\gamma m_p^2}\int_*^c\frac{V(\chi)}{V'\left(\chi\right)}\textrm{d}\chi\, .
\end{equation}

With the potential in this form our approach is essentially the same as the one taken by Wang in \cite{Wang:2010si} (see also \footnote{Wang defines $N=-\frac{1}{m_p^2}\int_*^c\frac{W-Q}{W_{\phi}}\textrm{d}\phi-\frac{1}{m_p^2}\int_*^c\frac{Q}{W_{\chi}}\textrm{d}\chi$, with $Q$ a suitably chosen function of $\phi$ and $\chi$ such that integrands are only functions of $\phi$ or $\chi$ as required.}). The case with $\alpha=0$ has to be treated separately and the potential takes the exponential form:
\begin{equation}\label{exppot}
W=W_0\,\textrm{Exp}\left[U(\phi)+V(\chi)\right]\, ,
\end{equation}
with $N$ given by (the factor of $1/2$ comes from the normalization of the exponent):
\begin{equation}\label{Nexpo}
N=-\frac{1}{2m_p^2}\int_*^c\frac{1}{U'\left(\phi\right)}\textrm{d}\phi
-\frac{1}{2m_p^2}\int_*^c\frac{1}{V'\left(\chi\right)}\textrm{d}\chi\, .
\end{equation}
In fact one could just write this as:
\begin{equation}\label{Nexpo2}
N=-\frac{1}{m_p^2}\int_*^c\frac{1}{U'\left(\phi\right)}\textrm{d}\phi\, ,
\end{equation}
with a similar expression for $\chi$. We keep the symmetric form for ready comparison with the homogeneous case.

\subsubsection*{The End of Inflation}
In addition to the restrictions on the potential (and possibly the trajectory) we also want the inflationary period to end with a phase of single field inflation in order to damp out the isocurvature perturbations and leave an adiabatic spectrum of fluctuations. During such a phase the isocurvature perturbations must become heavy while the adiabatic ones stay light. The relevant parameters are the effective masses of the adiabatic and isocurvature perturbations, given respectively by $\eta^{\sigma\sigma}$ and $\eta^{ss}$ (where $\sigma$ and $s$ are the directions parallel and perpendicular to the inflaton motion) \cite{Gordon:2000hv}:
\begin{align}\label{etass1}
\eta^{\sigma\sigma}&\equiv\frac{\epsilon^{\phi}\eta^{\phi}+2\sqrt{\epsilon^{\phi}\epsilon^{\chi}}\eta^{\phi\chi}+\epsilon^{\chi}\eta^{\chi}}{\epsilon}\nonumber\\
\eta^{ss}&\equiv\frac{\epsilon^{\chi}\eta^{\phi}-2\sqrt{\epsilon^{\phi}\epsilon^{\chi}}\eta^{\phi\chi}+\epsilon^{\phi}\eta^{\chi}}{\epsilon} \, .
\end{align}
When studying the fate of $f_{\mathrm{NL}}$ in section \ref{fnlfate} we will study what happens when $\eta^{ss}$ becomes large, while $\eta^{\sigma\sigma}$ stays small.

\subsubsection*{The Inflaton Trajectory and Violating Slow Roll}
Before moving on to our results, let us, for completeness, briefly consider what happens if trajectories deviate from slow roll. To derive the above the results (and those that follow below) we have worked in the slow roll regime. However, even if we insist on slow roll at horizon exit and as we approach the end of inflation, it is easy to imagine that slow roll may be violated briefly during either a sharp turn in field space (turning being both a hallmark of multifield inflaton \cite{Lyth:1998xn} and necessary for superhorizon generation of non-gaussianity \cite{Bernardeau:2002jy}) or during the approach to adiabaticity. To this end, let us consider what sort of violations of slow roll may be still tolerated in our framework.  For a different approach which utilizes the Hamilton-Jacobi formalism to study the evolution of the Hubble parameter without relying on slow roll, see \cite{Salopek:1990jq,Battefeld:2009ym,Byrnes:2009qy}.

To begin with let us consider $N$. We stipulate that slow roll must be imposed at horizon crossing ($t=t_{*}$) and at comoving surface on which we calculate our spectra ($t=t_c$). Then we consider how the derivatives of $N$ change if we violate slow roll in some region $t_1<t<t_2$, where $t_*<t_1<t_2<t_c$. Then, writing
\begin{align}
N&=\int_*^1H\textrm{d}t+\int_1^2H\textrm{d}t+\int_2^cH\textrm{d}t\, ,
\end{align}
one can show, with a little work and using the fact that slow roll conditions apply at $t_1$ and $t_2$ (but not between them):
\begin{align}
\frac{\partial N}{\partial\phi_*}&=\left.\frac{\partial N}{\partial\phi_*}\right|_{SR}+\int_1^2\frac{\partial}{\partial\phi_*}\left(\frac{H}{\dot{\phi}}\right)\textrm{d}\phi\, .
\end{align}
Here, $SR$ denotes the slow roll result (given in appendix \ref{details}). There is a similar result for $\partial N/\partial\chi_*$.

For $C$, we need to check under what circumstances it remains a good constant of motion outside the slow roll regime. To this end, consider the time derivative of $C$ without imposing slow roll. From (\ref{Cpotential}):
\begin{align}
\frac{\textrm{d}C}{\textrm{d}t}&=-m_p^2\frac{\dot{\phi}}{U'}+m_p^2\frac{\dot{\chi}}{V'}\nonumber\\
&=m_p^2\frac{F'U'+\ddot{\phi}}{3HU'}-m_p^2\frac{F'V'+\ddot{\chi}}{3HV'}\nonumber\\
&=m_p^2\frac{\ddot{\phi}}{3HU'}-m_p^2\frac{\ddot{\chi}}{3HV'}\, .
\end{align}
Thus for our framework to be valid we must satisfy the following conditions:
\begin{subequations}
\label{notSR}
\begin{align}
\int_1^2\left[\frac{\partial}{\partial\phi_*}\frac{H}{\dot{\phi}}\right]\textrm{d}\phi&\ll1\, ,\label{NphinotSR}\\
\int_1^2\left[\frac{\partial}{\partial\chi_*}\frac{H}{\dot{\chi}}\right]\textrm{d}\chi&\ll1\, ,\label{NchinotSR}\\
m_p^2\frac{\ddot{\phi}}{3HU'}-m_p^2\frac{\ddot{\chi}}{3HV'}&\ll1\, .\label{CnotSR}
\end{align}
\end{subequations}

The question we should ask, then, is: To what degree can we soften the slow roll criteria ($|\dot{H}|\ll H^2$ and $|\ddot{\phi_a}|\ll H|\dot{\phi_a}|$) and still satisfy the criteria of (\ref{notSR})? If we rewrite (\ref{NphinotSR}) and (\ref{NchinotSR}) so the integral is over the number of e-foldings, we have:
\begin{equation}
\int_1^2\frac{\dot{\phi}}{H}\left[\frac{\partial}{\partial\phi_*}\frac{H}{\dot{\phi}}\right]\textrm{d}N\ll1\, ,
\end{equation}
with a similar equation for $\chi$. From this we can immediately see that any sufficiently short violations of slow roll will, as one would expect, leave the equations for derivatives of $N$ unchanged.  This is, of course, unsurprising. However, the inequality can also be satisfied if the quantity $H/\dot{\phi}$ has only a weak dependance on $\phi_*$ compared with its magnitude. This latter requirement can be met if the trajectory before entering the region of slow roll violations is sufficiently attractive, so that $\dot{\phi}$ and $H$ are nearly independent of the initial field value. Unfortunately, unlike single field inflation, attractor behavior is not universal in two field inflation, so we can make no precise statements about our model that are potential and trajectory independent.

With regards to (\ref{CnotSR}), the most straightforward way for violations of slow roll to leave our results unchanged is, again, for said violation to be suitable short in duration. Alternatively the right hand side of (\ref{CnotSR}) will be zero if the friction term in (\ref{eqm}) is sub-dominant so that $\ddot{\phi}/U'=\ddot{\chi}/V'=F'$.

To summarize, then, violations of slow roll during the inflaton motion are consistent with our $\delta N$ analysis so long as either the violation is sufficiently short (in e-folding time) or if $H/\dot{\phi}$ is only weakly-dependent on the initial field value (attractive trajectory) and friction is subdominant during any non-slow roll regime. While it is unclear how general these scenarios are, we close this section by noting that there is some numerical evidence to lead us to believe that we can go beyond slow roll. For example, Vernizzi and Wands compare analytic and numerical results for doubly quadratic inflation in \cite{Vernizzi:2006ve} (see also \cite{Wanatanabe:2009}), for a variety of different initial conditions, and they find close agreement.

\subsection{Results}\label{results}
We will present our results for both homogeneous and exponential potentials here; the involved details of the calculation can be found in appendix \ref{details}, but the basic procedure is to take derivatives of $N$ and use the derivatives of $C$ to relate variations of $\phi_*$ to variations in $\phi_c$ (and likewise for $\chi$). As mentioned above have also studied the general potential $W(\phi,\chi)=F(U(\phi)+V(\chi))$, the details can be found in appendix \ref{limitedtraj}.

\subsubsection*{Homogeneous Potential: $W(\phi,\chi)=\left[U(\phi)+V(\chi)\right]^{\gamma}$}
Results here are expressed in terms of derivatives of the potential (as before a prime denotes a derivative with respect to the argument of the function -- $\phi$ for $U$, $\chi$ for $V$ and $U+V$ for $F$) and the slow roll parameters defined in (\ref{slowroll}). First, the mass parameters for the adiabatic and isocurvature fluctuations are given by:
\begin{align}\label{etass2}
\eta^{\sigma\sigma}&=\frac{\epsilon^{\phi}\eta^{\phi}+4\frac{(\gamma-1)}{\gamma}\epsilon^{\phi}\epsilon^{\chi}+\epsilon^{\chi}\eta^{\chi}}{\epsilon}\, ,\nonumber\\
\eta^{ss}&=\frac{\epsilon^{\chi}\eta^{\phi}-4\frac{(\gamma-1)}{\gamma}\epsilon^{\phi}\epsilon^{\chi}+\epsilon^{\phi}\eta^{\chi}}{\epsilon}\, .
\end{align}

For the other quantities of interest, we first define:
\begin{align}\label{xy}
x_h&\equiv\frac{1}{U_*+V_*}\left(U_*+\frac{V_c\epsilon_c^{\phi}-U_c\epsilon_c^{\chi}}{\epsilon_c}\right)\, ,\nonumber\\
y_h&\equiv\frac{1}{U_*+V_*}\left(V_*-\frac{V_c\epsilon_c^{\phi}-U_c\epsilon_c^{\chi}}{\epsilon_c}\right)\, .
\end{align}
Then the observables $\mathcal{P}_{\zeta}$ and $n_{\zeta}-1$ are given by:
\begin{widetext}
\begin{align}\label{Pandetahomogeneous}
\mathcal{P}_{\zeta}&=\frac{W_*}{24\pi^2m_p^4}\left(\frac{x_h^2}{\epsilon_*^{\phi}}+\frac{y_h^2}{\epsilon_*^{\chi}}\right)\, ,\nonumber\\
n_{\zeta}-1&=-2\epsilon_*-\frac{4}{\gamma}\left(\frac{x_h\left[1-\left(\frac{\gamma\eta^{\phi}_*}{2\epsilon_*^{\phi}}-\gamma+1\right)x_h\right]+y_h\left[1-\left(\frac{\gamma\eta^{\chi}_*}{2\epsilon_*^{\chi}}-\gamma+1\right)y_h\right]}{\frac{x_h^2}{\epsilon_*^{\phi}}+\frac{y_h^2}{\epsilon_*^{\chi}}}\right)\, ,
\end{align}
and $f_{\mathrm{NL}}^{(4)}$ by:
\begin{multline}\label{fnlhomogeneous}
\frac{6}{5}f_{\mathrm{NL}}^{(4)}= \frac{2}{\gamma}\left(\frac{\frac{x_h^2}{\epsilon^{\phi}_*}\left[1-\left(\frac{\gamma\eta^{\phi}_*}{2\epsilon_*^{\phi}}-\gamma+1\right)x_h\right]+\frac{y_h^2}{\epsilon^{\chi}_*}\left[1-\left(\frac{\gamma\eta^{\chi}_*}{2\epsilon_*^{\chi}}-\gamma+1\right)y_h\right]}{\left(\frac{x_h^2}{\epsilon_*^{\phi}}+\frac{y_h^2}{\epsilon_*^{\chi}}\right)^2}\right)\\
+\frac{2}{\gamma}\left(\frac{\frac{\left(U_c+V_c\right)^2}{\left(U_*+V_*\right)^2}\left(\frac{x_h}{\epsilon_*^{\phi}}-\frac{y_h}{\epsilon_*^{\chi}}\right)^2\frac{\epsilon_c^{\phi}\epsilon_c^{\chi}}{\epsilon_c}\left(\frac{\gamma\eta_c^{ss}}{\epsilon_c}-1\right)}{\left(\frac{x_h^2}{\epsilon_*^{\phi}}+\frac{y_h^2}{\epsilon_*^{\chi}}\right)^2}\right)\, .
\end{multline}
\end{widetext}

\subsubsection*{Exponential Potential: $W(\phi,\chi)=W_0\textrm{Exp}\left[U(\phi)+V(\chi)\right]$}
The slow roll parameters (\ref{slowroll}) take a particularly simple form for an exponential potential of the form of (\ref{exppot}), and readily lead to the following expressions for the mass parameters $\eta^{ss}$ and $\eta^{\sigma\sigma}$:
\begin{align}\label{etassexp}
\eta^{\sigma\sigma}&\equiv\frac{\epsilon^{\phi}\eta^{\phi}+4\epsilon^{\phi}\epsilon^{\chi}+\epsilon^{\chi}\eta^{\chi}}{\epsilon}\, ,\nonumber\\
\eta^{ss}&\equiv\frac{\epsilon^{\chi}\eta^{\phi}-4\epsilon^{\phi}\epsilon^{\chi}+\epsilon^{\phi}\eta^{\chi}}{\epsilon} \, .
\end{align}

Next, in a similar fashion to the previous section, we define:
\begin{align}\label{xye}
x_e&\equiv 1+\frac{\epsilon_c^{\phi}-\epsilon_c^{\chi}}{\epsilon_c}\, , & y_e&\equiv 1-\frac{\epsilon_c^{\phi}-\epsilon_c^{\chi}}{\epsilon_c}\, .
\end{align}
Then the observables $\mathcal{P}_{\zeta}$ and $n_{\zeta}-1$ are given by:
\begin{align}\label{Pandnsexp}
\mathcal{P}_{\zeta}&=\frac{W_*}{96\pi^2m_p^4}\left(\frac{x_e^2}{\epsilon_*^{\phi}}+\frac{y_e^2}{\epsilon_*^{\chi}}\right)\, ,\nonumber\\
n_{\zeta}-1&=-2\epsilon_*-4\left(\frac{2-\left(\frac{\eta_*^{\phi}x_e^2}{2\epsilon_*^{\phi}}+2x_ey_e+\frac{\eta_*^{\chi}y_e^2}{2\epsilon_*^{\chi}}\right)}{\frac{x_e^2}{\epsilon_*^{\phi}}+\frac{y_e^2}{\epsilon_*^{\chi}}}\right)\, ,
\end{align}
and $\frac{6}{5}f_{\mathrm{NL}}^{(4)}$ by:
\begin{equation}
\label{fnlexp}
\frac{1}{2}\frac{\frac{4\epsilon_c^{\phi}\epsilon_c^{\chi}}{\epsilon_c^2}\eta_c^{ss}\left(\frac{x_e}{\epsilon_*^{\phi}}-\frac{y_e}{\epsilon_*^{\chi}}\right)^2-x_e^3\left(\frac{\eta_*^{\phi}-2\epsilon_*^{\phi}}{\left(\epsilon_*^{\phi}\right)^2}\right)-y_e^3\left(\frac{\eta_*^{\chi}-2\epsilon_*^{\chi}}{\left(\epsilon_*^{\chi}\right)^2}\right)}{\left(\frac{x_e^2}{2\epsilon_*^{\phi}}+\frac{y_e^2}{2\epsilon_*^{\chi}}\right)^2}\, .
\end{equation}

\section{Damping Away Isocurvature and the Fate of $f_{\mathrm{NL}}$}\label{fnlfate}
\subsubsection*{Homogeneous Potential: $W(\phi,\chi)=\left[U(\phi)+V(\chi)\right]^{\gamma}$}
We are now in a position to assess the magnitude of $f_{\mathrm{NL}}^{(4)}$, beginning with the homogeneous potential. Following the arguments of \cite{Vernizzi:2006ve}, we see that $x_h+y_h=1$ and max$(x_h,y_h)\le1$, where $x_h$ and $y_h$ are defined in (\ref{xy}), so the denominator of (\ref{fnlhomogeneous}) is of order $\varepsilon_*^{-2}$, where $\varepsilon$ refers to a generic first-order slow roll parameter.  Both terms in the numerator in the first line of (\ref{fnlhomogeneous}) are of order $\varepsilon_*^{-1}$, so the first term in parentheses is of order $\varepsilon_*$.  The parameter $\gamma$ which appears in the denominator of (\ref{fnlhomogeneous}) is in principle unconstrained, so it might be expected that choosing $\gamma$ to be smaller than $\varepsilon_*$ could produce large non-gaussianity.  However, a factor $\gamma^{-1}$ also appears in (\ref{Pandetahomogeneous}) for the scalar spectral index $n_{\zeta}-1$.  Observation of a nearly scale-invariant power spectrum \cite{Komatsu:2010fb} thus requires that $|\gamma^{-1}| \lesssim \mathcal{O}(1)$.  Returning to (\ref{fnlhomogeneous}), we thus see that the second line gives the only contribution to $f_{\mathrm{NL}}^{(4)}$ which is not automatically slow roll suppressed:
\begin{align}\label{fnldom}
\frac{6}{5}f_{\mathrm{NL}}^{(4)}&\sim\mathcal{O}(\varepsilon_*)+2\frac{\frac{\left(U_c+V_c\right)^2}{\left(U_*+V_*\right)^2}\left(\frac{x_h}{\epsilon_*^{\phi}}-\frac{y_h}{\epsilon_*^{\chi}}\right)^2}{\left(\frac{x_h^2}{\epsilon_*^{\phi}}+\frac{y_h^2}{\epsilon_*^{\chi}}\right)^2}\frac{\epsilon_c^{\phi}\epsilon_c^{\chi}}{\epsilon_c^2}\eta_c^{ss}\nonumber\\
&\sim\mathcal{O}(\varepsilon_*)+\mathcal{O}(1)\times\frac{\epsilon_c^{\phi}\epsilon_c^{\chi}}{\epsilon_c^2}\eta_c^{ss}\, .
\end{align}

We are interested in the value of $f_{\mathrm{NL}}^{(4)}$ only after it becomes a conserved quantity -- while there may large non-gaussianities at any given time during inflation, $f_{\mathrm{NL}}^{(4)}$ is a dynamic quantity and we must follow its evolution until it becomes constant and we can readily translate the primordial value into observables today. Alternatively, one could also follow the evolution up to the time of observation itself, but that would require detailed knowledge of the cosmological history from the present all the way back to the inflationary epoch. Given our ignorance of the details of much of the early universe, we focus on the former approach. Accordingly we wish to study $f_{\mathrm{NL}}^{(4)}$ as it is forced into a constant value and, since correlations of the curvature perturbation will evolve outside the horizon as long as non-adiabatic fluctuations are present, this means we need to evaluate (\ref{fnldom}) after any non-adiabaticity has been damped away and only the adiabatic mode is left.

As discussed above, we are going to damp away non-adiabatic fluctuations by passing through a phase of effectively single field inflation, which requires that $\eta^{ss}$ becomes large during a finite period before the end of inflation.  Now, in the slow roll approximation the equations of motion for the adiabatic and isocurvature fluctuations decouple \cite{Gordon:2000hv}, and we have for the latter:
\begin{equation}\label{nonadiabatic}
\ddot{\delta s} + 3H\dot{\delta s} + \frac{W}{m_P^2}\eta^{ss}\delta s = 0.
\end{equation}
The solution of this equation in the slow roll approximation is \cite{Gordon:2000hv,Riotto:2002yw}
\begin{equation}\label{deltasgeneral}
\delta s \propto a(t)^{-3/2}\left(\frac{k}{aH}\right)^{-\nu}
\end{equation}
where $\nu$ is given by
\begin{equation}\label{nu}
\nu^2=\frac{9}{4}-\frac{W\eta^{ss}}{m_P^2H^2}.
\end{equation}	
So for $\eta^{ss} \geq \frac{3}{4}$, we find
\begin{equation}\label{deltas}
|\delta s| \propto a(t)^{-3/2}
\end{equation}
and the isocurvature fluctuations are rapidly damped away.

We will now examine the conditions for large $\eta^{ss}$ (defined in (\ref{etass1})). Recall that (\ref{slowroll}) gives
\begin{equation}\label{etaphichi}
\eta^{\phi\chi}=2\frac{(\gamma-1)}{\gamma}\sqrt{\epsilon^{\phi}\epsilon^{\chi}}
\end{equation}
and $|\gamma|$ must be $\mathcal{O}(1)$ or larger to guarantee scale invariance.  Thus $\eta^{\phi\chi}$ cannot give the dominant contribution to $\eta^{ss}$ when
\begin{equation}\label{etass3}
\eta^{ss}=\frac{\epsilon^{\chi}\eta^{\phi}-4\frac{(\gamma-1)}{\gamma}\epsilon^{\phi}\epsilon^{\chi}+\epsilon^{\phi}\eta^{\chi}}{\epsilon}
\end{equation}
is large.  As a result, we must have either $\eta^{\phi}\gtrsim\frac{\epsilon}{\epsilon^{\chi}}$ or $\eta^{\chi}\gtrsim\frac{\epsilon}{\epsilon^{\phi}}$.  If both $\eta^{\phi}$ and $\eta^{\chi}$  are large, then $\eta^{\sigma\sigma}$ will also be large, and inflation will quickly end before the non-adiabatic modes have been damped away, so we will not be interested in this case.

Because $\eta^{ss}$ appears in the second line of (\ref{fnlhomogeneous}), it might be expected that for sufficiently large $\eta^{ss}$, observably large non-gaussianity is produced in the process of damping away non-adiabatic fluctuations.  However, the coefficient of $\eta^{ss}$, specifically $\frac{\epsilon^{\phi}\epsilon^{\chi}}{\epsilon^2}$, is a dynamical quantity whose time-dependence is affected by the behavior of $\eta^{ss}$.  To find this time-dependence, we first we note that:
\begin{align}
\frac{\dot{\epsilon^\phi}}{H} &=-2\epsilon^\phi\eta^\phi-2\sqrt{\epsilon^\phi\epsilon^\chi}\eta^{\phi\chi}+4(\epsilon^\phi)^2+4\epsilon^\phi\epsilon^\chi \label{epsilonphidot}\\
\frac{\dot{\epsilon^\chi}}{H} &=-2\epsilon^\chi\eta^\chi-2\sqrt{\epsilon^\phi\epsilon^\chi}\eta^{\phi\chi}+4(\epsilon^\chi)^2+4\epsilon^\phi\epsilon^\chi \label{epsilonchidot}.
\end{align}
Now, if we have $\eta^{\phi}>1$, we can neglect all but the first term on the right hand side of (\ref{epsilonphidot}), and so we find that
\begin{equation}\label{epsilonphidamping}
\epsilon^{\phi}(t) \propto \textrm{Exp}\left[-2\int H\eta^{\phi}\rm{d}t\right].
\end{equation}
Similar remarks apply to $\epsilon^{\chi}$, in the case that $\eta^{\chi}>1$.

As a result, we find that after $\eta^{ss}$ becomes large, the dominant contribution to $f_{\mathrm{NL}}^{(4)}$ decays exponentially with time, with a decay constant greater than that of the isocurvature fluctuations.  Specifically, from (\ref{fnldom}) and (\ref{deltas}) we have for $\eta^{ss}>1$:
\begin{align}\label{damping}
|\delta s|&\sim \textrm{Exp}\left[-\frac{3}{2}\int H\textrm{d}t\right] \nonumber \\
f_{\mathrm{NL}}^{(4)}&\sim\mathcal{O}(\varepsilon_*)+\mathcal{O}(1)\times\eta^{ss}\textrm{Exp}\left[-2\int C_{\eta}H\eta^{ss}\textrm{d}t\right]\, .
\end{align}
$C_{\eta}$ is a number which is always greater than 1 whose value depends on the particular direction of the effective single field during this phase. We thus conclude that $f_{\mathrm{NL}}^{(4)}$ will always be slow roll suppressed upon entering the purely adiabatic solution after a phase of effectively single field inflation.

There are at least two cases in which the terms labeled $\mathcal{O}(\varepsilon_*)$ may be significant, even though those terms are proportional to slow roll parameters.  In the first case, these terms are initially small, but may become large for finely tuned trajectories in field space \cite{Byrnes:2008wi}.  In this case, these terms will always be small after passing through a phase of effectively single field inflation for reasons similar to those discussed above.  In the second case, the individual terms contained in those labeled $\mathcal{O}(\varepsilon_*)$ are significant, but there is initially a cancellation so that $f_{\mathrm{NL}}^{(4)}$ is small at horizon exit \cite{Kim:2010ud} (the model studied in \cite{Kim:2010ud} has a large number of fields, but a similar mechanism should work with two fields and an appropriately chosen potential).  After passing through a phase of effectively single field inflation, various terms are damped, and there is no longer a cancellation near the end of inflation.  This has the possibility for giving a significant (though not parametrically large) $f_{\mathrm{NL}}$ and purely adiabatic fluctuations at the end of inflation.

\subsubsection*{Exponential Potential: $W(\phi,\chi)=W_0\textrm{Exp}\left[U(\phi)+V(\chi)\right]$}
The arguments above can be readily adapted to the exponential potential. By a similar argument as above, the dominant contribution to $f_{\mathrm{NL}}^{(4)}$ is proportional to $\eta^{ss}$:
\begin{align}\label{fnldomexpo}
\frac{6}{5}f_{\mathrm{NL}}^{(4)}&\sim\mathcal{O}(\varepsilon_*)+2\frac{\left(\frac{x_h}{\epsilon_*^{\phi}}-\frac{y_h}{\epsilon_*^{\chi}}\right)^2}{\left(\frac{x_h^2}{\epsilon_*^{\phi}}+\frac{y_h^2}{\epsilon_*^{\chi}}\right)^2}\frac{\epsilon_c^{\phi}\epsilon_c^{\chi}}{\epsilon_c^2}\eta_c^{ss}\nonumber\\
&\sim\mathcal{O}(\varepsilon_*)+\mathcal{O}(1)\times\frac{\epsilon_c^{\phi}\epsilon_c^{\chi}}{\epsilon_c^2}\eta_c^{ss}\, .
\end{align}
Once again, in order to damp out non-adiabatic fluctuations, we need large $\eta^{ss}$, which in turn implies either large $\eta^{\phi}$ or large $\eta^{\chi}$:
\begin{align}
\eta^{ss}&\equiv\frac{\epsilon^{\chi}\eta^{\phi}-4\epsilon^{\phi}\epsilon^{\chi}+\epsilon^{\phi}\eta^{\chi}}{\epsilon} \, .
\end{align}
The rest of the analysis is identical to the previous section and once again leads to exponential damping of $f_{\mathrm{NL}}^{(4)}$ of the form given in (\ref{damping}).

\subsubsection*{The General Case}
A similar argument (though somewhat more involved and with a significant loss in clarity as payment for the increased generality) holds for the general potential $W(\phi,\chi)=F(U(\phi)+V(\chi))$, and is presented in appendix \ref{limitedtraj}.

\section{More General Models and other Approaches to Adiabaticity}\label{evading}
Although we have shown that $f_{\mathrm{NL}}^{\textrm{local}}$ is suppressed during the approach to adiabaticity in a particular class of two field inflationary models, it would be helpful to understand how generally this conclusion applies.  The importance of this question is made clear when we recall that as long as non-adiabatic fluctuations are present, the curvature perturbation and its correlation functions will evolve outside the horizon. Therefore no sharp predictions can be made without an understanding of the evolution up to the time of adiabaticity or observation. Consequently, no model that relies upon the superhorizon evolution of the curvature perturbation to generate non-gaussianity can claim to predict a large value for $f_{\mathrm{NL}}^{\textrm{local}}$ without a calculation which determines its evolution up to adiabaticity or observation.

There are two broad possibilities for generalizing the above results: changing the field dynamics or changing the mechanism for achieving adiabaticity.

\subsubsection*{Dynamics}
To change the dynamics one can either look at a wider set of trajectories, consider a larger class of potentials, or extend the number of inflaton fields. We have already discussed the applicability of our results if the inflaton motion violates slow roll at the end of section \ref{potentialdetails}. While we see no reason why similar results to those derived above should not hold for more general potentials, we have not proved the applicability of our conclusions beyond the class of potentials that we have studied. What's more, as detailed in section \ref{potentialdetails}, the particular potential we consider is, in some sense, the most general that can be readily studied analytically using the $\delta N$ formalism. Accordingly, to explore a larger set of potentials we would first need to develop alternative (efficient) ways of extracting the statistical parameters of interest (the power spectrum, the bispectrum etc.).

For more than two fields, we expect the result to hold for appropriately slow roll trajectories in potentials of the form $W(\phi_1,\ldots,\phi_n)=F(U_1(\phi_1)+\ldots+U_n(\phi_n))$, because in this case we can construct a set of conserved quantities \cite{Battefeld:2006sz} $C_1,\ldots,C_{n-1}$ by the same method as explained in section \ref{potentialdetails}. For more general potentials and trajectories, no definitive statement can be made.

\subsubsection*{The Approach to Adiabaticity}
Instead of using a phase of single field inflation to drive the cosmological fluctuations into an adiabatic solution, one can instead achieve the same result if the universe goes through a period of local thermal equilibrium with no non-zero conserved quantum numbers \cite{Weinberg:2004kf,Weinberg:2008si}.  An understanding of this process would allow a fuller treatment of models in which non-adiabatic fluctuations persist through the end of inflation including more general models of multiple field inflation, modulated reheating, and the curvaton scenario.  We plan to explore this issue in future work, and while there is no definitive answer yet, we believe that local thermal equilibrium should have a similar damping effect as a phase of single field inflation.

Essentially, this is because the relevant underlying dynamics during a period of local thermal equilibrium are very similar to those during a phase of single field inflation: there is single scalar degree of the freedom (either the inflaton or the temperature) which in turn means that there are only two solutions for the perturbations in the cosmological fluid, both of which must be adiabatic \cite{Weinberg:2004kf}. Moreover, the attractiveness of the adiabatic solution as one approaches thermal equilibrium is at least qualitatively similar to what happens as one approaches a phase of single field inflation. Of course, to show the damping (or otherwise) of $f_{\mathrm{NL}}$ one needs a model of reheating and thermalization to which one can readily apply either an analytical method such as the $\delta N$ formalism or a suitable numerical analysis.

\section{Conclusions}\label{conclusion}
In this paper, we have considered the generation of $f_{\mathrm{NL}}^{\textrm{local}}$ from superhorizon evolution of the curvature perturbation in two field inflation, in the most general class of potentials ($W(\phi,\chi)=F(U(\phi)+V(\chi))$) that seem readily tractable in the $\delta N$ formalism. Furthermore, we have shown that if the universe passes through a sufficiently long period of effectively single field inflation to force the cosmological perturbations into an adiabatic solution -- as suggested by observation and as is necessary to ensure subsequent conservation of the curvature perturbation -- $|f_{\mathrm{NL}}^{\textrm{local}}|$ will be exponentially damped toward a value which is slow roll suppressed.

We stress, again, that predictions of an observable $f_{\mathrm{NL}}$ (or any other statistical property of the spectrum) from primordial fluctuations are only possible if one knows how the fluctuations evolve from the very early universe to today. Moreover, due to our ignorance of several phases of the early universe, the ability to make sharp predictions is restricted to the cases where there exists a conservation law which prevents some unknown dynamics from changing the statistical properties of the fluctuations.  Luckily, there is such a conservation law when the fluctuations are adiabatic. As such, multifield inflationary models which generate large values for $f_{\mathrm{NL}}^{\textrm{local}}$ from the superhorizon dynamics of the curvature perturbation are incomplete without an understanding of the subsequent evolution of the spectrum until constancy is achieved.


To put our results in context, recall that the measurement of non-gaussianities of the local form has been considered to be a sharp probe of the number of inflaton fields as $f_{\mathrm{NL}}^{\textrm{local}}$ is small in single field inflation, but could apparently grow large in the presence of multiple inflatons. This work shows, however, that while it is indeed possible to generate a significant contribution to the bispectrum in multifield inflation, such a contribution is necessarily transitory if one wishes to eliminate non-adiabaticity by a phase of single field inflaton. Accordingly, a significant measurement of $f_{\mathrm{NL}}^{\textrm{local}}$ in the current generation of cosmological experiments would present a substantial challenge to inflationary model builders.

\subsection*{Note}
After the first version of this work appeared, a paper by Peterson and Tegmark \cite{Peterson:2010mv} appeared which shows similar results, using an alternative framework for analyzing the evolution of inflationary perturbations.

\begin{acknowledgments}
The authors would like to thank Christian Byrnes, Raphael Flauger, Eiichiro Komatsu, David Seery, and Steven Weinberg for helpful discussions. This material is based upon work supported by the National Science Foundation under Grant No. PHY-0455649 and by the Texas Cosmology Center, which is supported by the College of Natural Sciences and the Department of Astronomy at the University of Texas at Austin and the McDonald Observatory.  J.M. was also supported by the A. D. Hutchison Student Endowment Fellowship during the completion of this work.
\end{acknowledgments}

\appendix
\section{Details for Homogenous and Exponential Potentials}\label{details}
\subsection*{Homogeneous Potential: $W(\phi,\chi)=\left[U(\phi)+V(\phi)\right]^{\gamma}$}
With $N$ and $C$ given by (\ref{Cpotential}) and (\ref{Nhomogeneous}):
\begin{align}
C&=-m_p^2\int_{\phi_0}^{\phi}\frac{1}{U'(\phi')}\textrm{d}\,\phi'+m_p^2\int_{\chi_0}^{\chi}\frac{1}{U'(\chi')}\,\textrm{d}\chi'\tag{\ref{Cpotential}}\, ,\\
N&=-\frac{1}{\gamma m_p^2}\int_*^c\frac{U(\phi)}{U'\left(\phi\right)}\textrm{d}\phi-\frac{1}{\gamma m_p^2}\int_*^c\frac{V(\chi)}{V'\left(\chi\right)}\textrm{d}\chi\, . \tag{\ref{Nhomogeneous}}
\end{align}
Varying $N$ then gives:
\begin{multline}\label{dNhomogeneous}
\textrm{d}N=\frac{1}{m_p^2\gamma}\left[\left(\frac{U}{U'}\right)_*-\frac{\partial\phi_c}{\partial\phi_*}\left(\frac{U}{U'}\right)_c-\frac{\partial\chi_c}{\partial\phi_*}\left(\frac{V}{V'}\right)_c\right]\textrm{d}\phi_*\\
+\frac{1}{m_p^2\gamma}\left[\left(\frac{V}{V'}\right)_*-\frac{\partial\phi_c}{\partial\chi_*}\left(\frac{U}{U'}\right)_c-\frac{\partial\chi_c}{\partial\chi_*}\left(\frac{V}{V'}\right)_c\right]\textrm{d}\chi_*\, .
\end{multline}
Note that in deriving the above we had to account for the dependence of $\phi_c$ and $\chi_c$ on both $\phi_*$ and $\chi_*$. We will also need:
\begin{align}\label{ctostar}
\textrm{d}\phi_c&=\frac{\textrm{d}\phi_c}{\textrm{d}C}\left(\frac{\partial C}{\partial\phi_*}\textrm{d}\phi_*+\frac{\partial C}{\partial\chi_*}\textrm{d}\chi_*\right)\, ,\nonumber\\
\textrm{d}\chi_c&=\frac{\textrm{d}\chi_c}{\textrm{d}C}\left(\frac{\partial C}{\partial\phi_*}\textrm{d}\phi_*+\frac{\partial C}{\partial\chi_*}\textrm{d}\chi_*\right)\, .
\end{align}
From (\ref{Cpotential}) we have:
\begin{align}\label{dCdphistar}
\frac{\partial C}{\partial\phi_*}&=-\frac{m_p^2}{U'_*}\, , & \frac{\partial C}{\partial\chi_*}&=\frac{m_p^2}{V'_*}\, ,
\end{align}
and
\begin{align}
\frac{\partial C}{\partial\phi_c}&=-\frac{m_p^2}{U'_c}\, , & \frac{\partial C}{\partial\chi_c}&=\frac{m_p^2}{V'_c}\, .
\end{align}
The time $t_c$ defines a surface of constant energy:
\begin{equation}
W\left(\phi_c,\chi_c\right)=\textrm{constant}\, .
\end{equation}
Differentiating with respect to $C$ then gives:
\begin{equation}
\frac{\textrm{d}\phi_c}{\textrm{d}C}\left.W_{\phi}\right|_c+\frac{\textrm{d}\chi_c}{\textrm{d}C}\left.W_{\chi}\right|_c=0\, .
\end{equation}
Using the above and $W_{\phi}/W_{\chi}=U'/V'$, we can differentiate the expression for $C$ in (\ref{Cpotential}) and obtain (after some manipulation) the following expressions:
\begin{gather}
\frac{\textrm{d}\phi_c}{\textrm{d}C}=-\frac{1}{m_p^2}\left[U'_c\left(\frac{1}{U^{'2}_c}+\frac{1}{V^{'2}_c}\right)\right]^{-1}=-\frac{1}{m_p^2}\frac{U'_cV^{'2}_c}{U^{'2}_c+V^{'2}_c}\, ,\nonumber\\
\frac{\textrm{d}\chi_c}{\textrm{d}C}=\frac{1}{m_p^2}\left[V'_c\left(\frac{1}{U^{'2}_c}+\frac{1}{V^{'2}_c}\right)\right]^{-1}=\frac{1}{m_p^2}\frac{U^{'2}_cV'_c}{U^{'2}_c+V^{'2}_c}\, .\label{dphidCc}
\end{gather}
Substituting (\ref{dCdphistar}) and (\ref{dphidCc}) into (\ref{ctostar}) allows us to read off the following:
\begin{align}\label{dcdstar}
\frac{\partial\phi_c}{\partial\phi_*}&=\frac{V^{'2}_c}{U^{'2}_c+V^{'2}_c}\frac{U_c'}{U_*'}\, , & \frac{\partial\phi_c}{\partial\chi_*}&=-\frac{V^{'2}_c}{U^{'2}_c+V^{'2}_c}\frac{U_c'}{V_*'}\, ,\nonumber\\
\frac{\partial\chi_c}{\partial\phi_*}&=-\frac{U^{'2}_c}{U^{'2}_c+V^{'2}_c}\frac{V_c'}{U_*'}\, , & \frac{\partial\chi_c}{\partial\chi_*}&=\frac{U^{'2}_c}{U^{'2}_c+V^{'2}_c}\frac{V_c'}{V_*'}\, .
\end{align}
The derivatives of $N$ are then:
\begin{align}
\frac{\partial N}{\partial\phi_*}&=\frac{1}{m_p}\frac{x_h}{\sqrt{2\epsilon_*^{\phi}}}\, , & \frac{\partial N}{\partial\chi_*}&=\frac{1}{m_p}\frac{y_h}{\sqrt{2\epsilon_*^{\chi}}}\, .
\end{align}
We have used the slow roll parameters defined in (\ref{slowroll}); while $x_h$ and $y_h$ are defined in (\ref{xy}):
\begin{align}
x_h&\equiv\frac{1}{U_*+V_*}\left(U_*+\frac{V_c\epsilon_c^{\phi}-U_c\epsilon_c^{\chi}}{\epsilon_c}\right)\, ,\nonumber\\
y_h&\equiv\frac{1}{U_*+V_*}\left(V_*-\frac{V_c\epsilon_c^{\phi}-U_c\epsilon_c^{\chi}}{\epsilon_c}\right)\, . \tag{\ref{xy}}
\end{align}
In a similar vein, we can find the second derivatives:
\begin{align}
\begin{split}
\frac{\partial^2N}{\partial\phi_*^2}&=\frac{1}{\gamma m_p^2}\left[1-x_h\left(\frac{\gamma\eta_*^{\phi}}{2\epsilon_*^{\phi}}-\gamma+1\right)\right.\\
&\qquad\qquad\left.+\frac{\left(U_c+V_c\right)^2}{\left(U_*+V_*\right)^2}\frac{1}{\epsilon_*^{\phi}}\frac{\epsilon_c^{\phi}\epsilon_c^{\chi}}{\epsilon_c}\left(\frac{\gamma\eta^{ss}_c}{\epsilon_c}-1\right)\right]
\end{split}\, \nonumber\\
\begin{split}
\frac{\partial^2N}{\partial\chi_*^2}&=\frac{1}{\gamma m_p^2}\left[1-y_h\left(\frac{\gamma\eta_*^{\chi}}{2\epsilon_*^{\chi}}-\gamma+1\right)\right.\\
&\qquad\qquad\left.+\frac{\left(U_c+V_c\right)^2}{\left(U_*+V_*\right)^2}\frac{1}{\epsilon_*^{\chi}}\frac{\epsilon_c^{\phi}\epsilon_c^{\chi}}{\epsilon_c}\left(\frac{\gamma\eta^{ss}_c}{\epsilon_c}-1\right)\right]
\end{split}\, \nonumber\\
\frac{\partial^2N}{\partial\phi_*\chi_*}&=\frac{1}{\gamma m_p^2}\left[-\frac{\left(U_c+V_c\right)^2}{\left(U_*+V_*\right)^2}\frac{1}{\sqrt{\epsilon_*^{\phi}\epsilon_*^{\chi}}}\frac{\epsilon_c^{\phi}\epsilon_c^{\chi}}{\epsilon_c}\left(\frac{\gamma\eta^{ss}_c}{\epsilon_c}-1\right)\right]\, .
\end{align}
$\eta^{ss}$ is given in (\ref{etass2}):
\begin{align}
\eta^{ss}&=\frac{\epsilon^{\chi}\eta^{\phi}-4\frac{(\gamma-1)}{\gamma}\epsilon^{\phi}\epsilon^{\chi}+\epsilon^{\phi}\eta^{\chi}}{\epsilon}\, .\tag{\ref{etass2}}
\end{align}
From these it is straightforward to use the relevant $\delta N$ equations to obtain the results given in (\ref{Pandetahomogeneous}) and (\ref{fnlhomogeneous}).

\subsection*{Exponential Potential: $W(\phi,\chi)=W_0\textrm{Exp}\left[U(\phi)+V(\chi)\right]$}
With $N$ from (\ref{Nexpo}) and $C$ again given by (\ref{Cpotential}),
\begin{align}
C&=-m_p^2\int_{\phi_0}^{\phi}\frac{1}{U'(\phi')}\textrm{d}\,\phi'+m_p^2\int_{\chi_0}^{\chi}\frac{1}{U'(\chi')}\,\textrm{d}\chi'\tag{\ref{Cpotential}}\\
N&=-\frac{1}{2m_p^2}\int_*^c\frac{1}{U'\left(\phi\right)}\textrm{d}\phi
-\frac{1}{2m_p^2}\int_*^c\frac{1}{V'\left(\chi\right)}\textrm{d}\chi\, ,\tag{\ref{Nexpo}}
\end{align}
the variation gives:
\begin{multline}\label{dNexpo}
\textrm{d}N=\frac{1}{2m_p^2}\left[\left(\frac{1}{U'}\right)_*-\frac{\partial\phi_c}{\partial\phi_*}\left(\frac{1}{U'}\right)_c-\frac{\partial\chi_c}{\partial\phi_*}\left(\frac{1}{V'}\right)_c\right]\textrm{d}\phi_*\\
+\frac{1}{2m_p^2}\left[\left(\frac{1}{V'}\right)_*-\frac{\partial\phi_c}{\partial\chi_*}\left(\frac{1}{U'}\right)_c-\frac{\partial\chi_c}{\partial\chi_*}\left(\frac{1}{V'}\right)_c\right]\textrm{d}\chi_*\, .
\end{multline}
The analysis begins in the same fashion as in the previous section, with the expressions for the derivatives of $\phi_c$ and $\chi_c$ with respect to $\phi_*$ and $\chi_*$, given in (\ref{dcdstar}). From this and (\ref{dNexpo}) we obtain the following expressions for the first derivatives of $N$:
\begin{align}
\frac{\partial N}{\partial\phi_*}&=\frac{1}{2m_p}\frac{x_e}{\sqrt{2\epsilon_*^{\phi}}}\, , & \frac{\partial N}{\partial\chi_*}&=\frac{1}{2m_p}\frac{y_e}{\sqrt{2\epsilon_*^{\chi}}}\, .
\end{align}
The slow roll parameters are defined in (\ref{slowroll}). $x_e$ and $y_e$ are defined in (\ref{xye}) as:
\begin{align}
x_e&\equiv 1+\frac{\epsilon_c^{\phi}-\epsilon_c^{\chi}}{\epsilon_c}\, ,\nonumber\\
y_e&\equiv 1-\frac{\epsilon_c^{\phi}-\epsilon_c^{\chi}}{\epsilon_c}\, . \tag{\ref{xye}}
\end{align}
The second derivatives are then given by:
\begin{align}
\frac{\partial^2N}{\partial\phi_*^2}&=\frac{1}{2m_p^2}\frac{1}{2\epsilon_*^{\phi}}\left[-\left(\eta_*^{\phi}-2\epsilon_*^{\phi}\right)x_e+\frac{4\epsilon_c^{\phi}\epsilon_c^{\chi}}{\epsilon_c^2}\eta^{ss}_c\right]\, ,\nonumber\\
\frac{\partial^2N}{\partial\chi_*^2}&=\frac{1}{2m_p^2}\frac{1}{\sqrt{\epsilon_*^{\phi}\epsilon_*^{\chi}}}\frac{2\epsilon_c^{\phi}\epsilon_c^{\chi}}{\epsilon_c^2}\eta^{ss}_c\, ,\nonumber\\
\frac{\partial^2N}{\partial\phi_*\chi_*}&=\frac{1}{2m_p^2}\frac{1}{2\epsilon_*^{\chi}}\left[-\left(\eta_*^{\chi}-2\epsilon_*^{\chi}\right)y_e+\frac{4\epsilon_c^{\phi}\epsilon_c^{\chi}}{\epsilon_c^2}\eta^{ss}_c\right]\, .
\end{align}
$\eta^{ss}$ is given in (\ref{etassexp}):
\begin{align}
\eta^{\sigma\sigma}&\equiv\frac{\epsilon^{\phi}\eta^{\phi}+4\epsilon^{\phi}\epsilon^{\chi}+\epsilon^{\chi}\eta^{\chi}}{\epsilon}\, ,\nonumber\\
\eta^{ss}&\equiv\frac{\epsilon^{\chi}\eta^{\phi}-4\epsilon^{\phi}\epsilon^{\chi}+\epsilon^{\phi}\eta^{\chi}}{\epsilon} \, .\tag{\ref{etassexp}}
\end{align}
As above, the $\delta N$ equations then give the observables of (\ref{Pandnsexp}) and (\ref{fnlexp}).

\begin{widetext}
\section{General $W(\phi,\chi)=F\left(U(\phi)+V(\phi)\right)$ Results}\label{limitedtraj}
\subsection{$\delta N$ and Observables}
As before we begin with the equations for $C$ and $N$:
\begin{align}
C&=-m_p^2\int_{\phi_0}^{\phi}\frac{1}{U'(\phi')}\textrm{d}\,\phi'+m_p^2\int_{\chi_0}^{\chi}\frac{1}{U'(\chi')}\,\textrm{d}\chi'\, ,\nonumber\\
N&=-\frac{1}{2m_p^2}\int_{\phi_*}^{\phi_c}\frac{W\left(\phi,\chi(\phi)\right)}{W_{\phi}\left(\phi,\chi(\phi)\right)}\,\textrm{d}\phi
-\frac{1}{2m_p^2}\int_{\chi_*}^{\chi_c}\frac{W\left(\chi,\phi(\chi)\right)}{W_{\chi}\left(\chi,\phi(\chi)\right)}\,\textrm{d}\chi\, .\nonumber
\end{align}
Thus:
\begin{multline}\label{dNgen}
\textrm{d}N=\frac{1}{2m_p^2}\left[\left(\frac{W}{W_{\phi}}\right)_*-\frac{\partial\phi_c}{\partial\phi_*}\left(\frac{W}{W_{\phi}}\right)_c-I_1-\frac{\partial\chi_c}{\partial\phi_*}\left(\frac{W}{W_{\chi}}\right)_c-I_2\right]\textrm{d}\phi_*\\
+\frac{1}{2m_p^2}\left[\left(\frac{W}{W_{\chi}}\right)_*-\frac{\partial\phi_c}{\partial\chi_*}\left(\frac{W}{W_{\phi}}\right)_c-I_3-\frac{\partial\chi_c}{\partial\chi_*}\left(\frac{W}{W_{\chi}}\right)_c-I_4\right]\textrm{d}\chi_*\, .
\end{multline}
The quantities $I_1$, $I_2$, $I_3$ and $I_4$ are defined as:
\begin{align}
I_1&\equiv\int_*^c\left(\frac{W_{\chi}\left(\phi,\chi(\phi)\right)}{W_{\phi}\left(\phi,\chi(\phi)\right)}-\frac{W\left(\phi,\chi(\phi)\right)W_{\phi\chi}\left(\phi,\chi(\phi)\right)}{W_{\phi}^2\left(\phi,\chi(\phi)\right)}\right)\frac{\partial\chi(\phi)}{\partial\phi_*}\,\textrm{d}\phi\, ,\nonumber\\
I_2&\equiv\int_*^c\left(\frac{W_{\phi}\left(\phi(\chi),\phi\right)}{W_{\chi}\left(\phi(\chi),\chi\right)}-\frac{W\left(\phi(\chi),\chi\right)W_{\phi\chi}\left(\phi(\chi),\chi\right)}{W_{\chi}^2\left(\phi(\chi),\chi)\right)}\right)\frac{\partial\phi(\chi)}{\partial\phi_*}\,\textrm{d}\chi\, ,\nonumber\\
I_3&\equiv\int_*^c\left(\frac{W_{\chi}\left(\phi,\chi(\phi)\right)}{W_{\phi}\left(\phi,\chi(\phi)\right)}-\frac{W\left(\phi,\chi(\phi)\right)W_{\phi\chi}\left(\phi,\chi(\phi)\right)}{W_{\phi}^2\left(\phi,\chi(\phi)\right)}\right)\frac{\partial\chi(\phi)}{\partial\chi_*}\,\textrm{d}\phi\, ,\nonumber\\
I_4&\equiv\int_*^c\left(\frac{W_{\phi}\left(\phi(\chi),\phi\right)}{W_{\chi}\left(\phi(\chi),\chi\right)}-\frac{W\left(\phi(\chi),\chi\right)W_{\phi\chi}\left(\phi(\chi),\chi\right)}{W_{\chi}^2\left(\phi(\chi),\chi)\right)}\right)\frac{\partial\phi(\chi)}{\partial\chi_*}\,\textrm{d}\chi\, .
\end{align}

Armed with (\ref{dNgen}) and (\ref{dcdstar}) we can now find derivatives of $N$. For the first derivatives we have:
\begin{align}
\frac{\partial N}{\partial\phi_*}&=\frac{1}{2m_p^2}\left[\frac{m_p}{\sqrt{2\epsilon_*^{\phi}}}+\frac{m_p}{\sqrt{2\epsilon_*^{\phi}}}\frac{F_cF_*'}{F_c'F_*}\frac{\epsilon^{\phi}_c-\epsilon^{\chi}_c}{\epsilon_c}-I_1-I_2\right]\, ,\nonumber\\
\frac{\partial N}{\partial\chi_*}&=\frac{1}{2m_p^2}\left[\frac{m_p}{\sqrt{2\epsilon_*^{\chi}}}-\frac{m_p}{\sqrt{2\epsilon_*^{\chi}}}\frac{F_cF_*'}{F_c'F_*}\frac{\epsilon^{\phi}_c-\epsilon^{\chi}_c}{\epsilon_c}-I_3-I_4\right]\, .
\end{align}
We've used here (and below) the definitions of the slow roll parameters given in (\ref{slowroll}). For the second derivatives, things are a little more complicated:
\begin{align}
\frac{\partial^2N}{\partial\phi_*^2}&=\frac{1}{2m_p^2}\Bigg[1-\frac{\eta_*^{\phi}}{2\epsilon_*^{\phi}}-\frac{F_c}{F_c'}\left(\frac{F_*'}{F_*}\frac{\eta_*^{\phi}}{2\epsilon_*^{\phi}}-\frac{F_*''}{F_*'}\right)\frac{\epsilon^{\phi}_c-\epsilon^{\chi}_c}{\epsilon_c}+\frac{1}{\epsilon_*^{\phi}} \frac{\epsilon_c^{\phi}\epsilon_c^{\chi}}{\epsilon_c}\frac{F_cF_*'}{F_c'F_*}\left(1-\frac{F_cF_c''}{F_c'^2}+\frac{2\eta_c^{ss}}{\epsilon_c}\right)-J_1-J_3\Bigg]\, ,\nonumber\\
\frac{\partial^2N}{\partial\chi_*^2}&=\frac{1}{2m_p^2}\Bigg[1-\frac{\eta_*^{\chi}}{2\epsilon_*^{\chi}}+\frac{F_c}{F_c'}\left(\frac{F_*'}{F_*}\frac{\eta_*^{\chi}}{2\epsilon_*^{\chi}}-\frac{F_*''}{F_*'}\right)\frac{\epsilon^{\phi}_c-\epsilon^{\chi}_c}{\epsilon_c}+\frac{1}{\epsilon_*^{\chi}} \frac{\epsilon_c^{\phi}\epsilon_c^{\chi}}{\epsilon_c}\frac{F_cF_*'}{F_c'F_*}\left(1-\frac{F_cF_c''}{F_c'^2}+\frac{2\eta_c^{ss}}{\epsilon_c}\right)-J_6-J_8\Bigg]\, ,\nonumber\\
\frac{\partial^2N}{\partial\chi_*\partial\phi_*}&=\frac{1}{2m_p^2}\Bigg[\frac{\epsilon_*}{\sqrt{\epsilon_*^{\phi}\epsilon_*^{\chi}}}\left(1-\frac{F_*F_*''}{F^{'2}_*}\right)-\frac{1}{\sqrt{\epsilon_*^{\phi}\epsilon_*^{\chi}}}\frac{\epsilon_c^{\phi}\epsilon_c^{\chi}}{\epsilon_c}\frac{F_cF_*'}{F_c'F_*}\left(1-\frac{F_cF_c''}{F_c'^2}+\frac{2\eta_c^{ss}}{\epsilon_c}\right)-J_2-J_4\Bigg]\, .
\end{align}
To derive the above we've used the results:
\begin{align}
\frac{\partial I_1}{\partial\phi_*}+\frac{\partial I_2}{\partial\phi_*}&=-\left(1-\frac{F_cF_c''}{F_c^{'2}}\right)\frac{1}{U_*^{'2}}\frac{U_c^{'2}V_c^{'2}}{U_c^{'2}+V_c^{'2}}+J_1+J_3\, ,\nonumber\\
\frac{\partial I_1}{\partial\chi_*}+\frac{\partial I_2}{\partial\chi_*}&=-\frac{U_*'}{V_*'}\left(1-\frac{F_*F_*''}{F_*^{'2}}\right)+\left(1-\frac{F_cF_c''}{F_c^{'2}}\right)\frac{1}{U_*'V_*'}\frac{U_c^{'2}V_c^{'2}}{U_c^{'2}+V_c^{'2}}+J_2+J_4\, ,\nonumber\\
\frac{\partial I_3}{\partial\phi_*}+\frac{\partial I_4}{\partial\phi_*}&=-\frac{V_*'}{U_*'}\left(1-\frac{F_*F_*''}{F_*^{'2}}\right)+\left(1-\frac{F_cF_c''}{F_c^{'2}}\right)\frac{1}{U_*'V_*'}\frac{U_c^{'2}V_c^{'2}}{U_c^{'2}+V_c^{'2}}+J_5+J_7\, ,\nonumber\\
\frac{\partial I_3}{\partial\chi_*}+\frac{\partial I_4}{\partial\chi_*}&=-\left(1-\frac{F_cF_c''}{F_c^{'2}}\right)\frac{1}{V_*^{'2}}\frac{U_c^{'2}V_c^{'2}}{U_c^{'2}+V_c^{'2}}+J_6+J_8\, .
\end{align}
Along with the definitions:
\begin{align}
\begin{split}
J_1&\equiv\int_*^c\Bigg[\left(\frac{V''}{U'}\left(1-\frac{FF''}{F^{'2}}\right)+\frac{V^{'2}}{U'}\left(-\frac{F''}{F'}-\frac{FF''}{F^{'2}}+\frac{2FF^{''2}}{F^{'3}}\right)\right)\left(\frac{\partial\chi(\phi)}{\partial\phi_*}\right)^2\\
&\quad+\frac{V'}{U'}\left(1-\frac{FF''}{F^{'2}}\right)\frac{\partial^2\chi(\phi)}{\partial\phi_*^2}\Bigg]\textrm{d}\phi
\end{split}\, ,\nonumber\\
\begin{split}
J_2=J_5&\equiv\int_*^c\Bigg[\left(\frac{V''}{U'}\left(1-\frac{FF''}{F^{'2}}\right)+\frac{V^{'2}}{U'}\left(-\frac{F''}{F'}-\frac{FF''}{F^{'2}}+\frac{2FF^{''2}}{F^{'3}}\right)\right)\frac{\partial\chi(\phi)}{\partial\phi_*}\frac{\partial\chi(\phi)}{\partial\chi_*}\\
&\quad+\frac{V'}{U'}\left(1-\frac{FF''}{F^{'2}}\right)\frac{\partial^2\chi(\phi)}{\partial\phi_*\partial\chi_*}\Bigg]\textrm{d}\phi
\end{split}\, ,\nonumber\\
\begin{split}
J_3&\equiv\int_*^c\Bigg[\left(\frac{U''}{V'}\left(1-\frac{FF''}{F^{'2}}\right)+\frac{U^{'2}}{V'}\left(-\frac{F''}{F'}-\frac{FF''}{F^{'2}}+\frac{2FF^{''2}}{F^{'3}}\right)\right)\left(\frac{\partial\phi(\chi)}{\partial\phi_*}\right)^2\\
&\quad+\frac{U'}{V'}\left(1-\frac{FF''}{F^{'2}}\right)\frac{\partial^2\phi(\chi)}{\partial\phi_*^2}\Bigg]\textrm{d}\chi
\end{split}\, ,\nonumber\\
\begin{split}
J_4=J_7&\equiv\int_*^c\Bigg[\left(\frac{U''}{V'}\left(1-\frac{FF''}{F^{'2}}\right)+\frac{U^{'2}}{V'}\left(-\frac{F''}{F'}-\frac{FF''}{F^{'2}}+\frac{2FF^{''2}}{F^{'3}}\right)\right)\frac{\partial\phi(\chi)}{\partial\phi_*}\frac{\partial\phi(\chi)}{\partial\chi_*}\\
&\quad+\frac{U'}{V'}\left(1-\frac{FF''}{F^{'2}}\right)\frac{\partial^2\phi(\chi)}{\partial\phi_*\partial\chi_*}\Bigg]\textrm{d}\chi
\end{split}\, ,\nonumber\\
\begin{split}
J_6&\equiv\int_*^c\Bigg[\left(\frac{V''}{U'}\left(1-\frac{FF''}{F^{'2}}\right)+\frac{V^{'2}}{U'}\left(-\frac{F''}{F'}-\frac{FF''}{F^{'2}}+\frac{2FF^{''2}}{F^{'3}}\right)\right)\left(\frac{\partial\chi(\phi)}{\partial\chi_*}\right)^2\\
&\quad+\frac{V'}{U'}\left(1-\frac{FF''}{F^{'2}}\right)\frac{\partial^2\chi(\phi)}{\partial\chi_*^2}\Bigg]\textrm{d}\phi
\end{split}\, ,\nonumber\\
\begin{split}
J_8&\equiv\int_*^c\Bigg[\left(\frac{U''}{V'}\left(1-\frac{FF''}{F^{'2}}\right)+\frac{U^{'2}}{V'}\left(-\frac{F''}{F'}-\frac{FF''}{F^{'2}}+\frac{2FF^{''2}}{F^{'3}}\right)\right)\left(\frac{\partial\phi(\chi)}{\partial\chi_*}\right)^2\\
&\quad+\frac{U'}{V'}\left(1-\frac{FF''}{F^{'2}}\right)\frac{\partial^2\phi(\chi)}{\partial\chi_*^2}\Bigg]\textrm{d}\chi
\end{split}\, .
\end{align}
In these expressions functions of $\chi$ in $\phi$ integrals should be rewritten using the solutions $\chi(\phi)$ for the particular trajectory we are on, so that said integrals are well-defined (and vice-versa in the $\chi$ integrals). As mentioned in the main body of the text, this constrains us to look only at those trajectories where there is a one-to-one mapping between $\phi$ and $\chi$.

We can simplify these expressions slightly by defining, in analogy with the homogeneous and exponential cases above
\begin{align}\label{xf}
x_F &\equiv 1+\frac{F_cF_*'}{F_c'F_*}\frac{\epsilon_c^{\phi}-\epsilon_c^{\chi}}{\epsilon_c}\, ,\nonumber\\
y_F &\equiv 1-\frac{F_cF_*'}{F_c'F_*}\frac{\epsilon_c^{\phi}-\epsilon_c^{\chi}}{\epsilon_c}\, .
\end{align}

From the above expressions we can (with some work) use the $\delta N$ formalism to find expressions for $P_{\zeta}$, $n_{\zeta}$ and $f_{\mathrm{NL}}^{(4)}$ in terms of slow roll parameters and derivatives of the potential. First $P_{\zeta}$:
\begin{equation}	
P_{\zeta}=\frac{P_*}{8m_p^4}\Bigg[\frac{1}{\epsilon_*^{\phi}}\left(x_F-\frac{\sqrt{2\epsilon_*^{\phi}}}{m_P}(I_1+I_2)\right)^2 + \frac{1}{\epsilon_*^{\chi}}\left(y_F-\frac{\sqrt{2\epsilon_*^{\chi}}}{m_P}(I_3+I_4)\right)^2\Bigg]\, .
\end{equation}
Then, for $n_{\zeta}$:
\begin{align}\label{nsF}
\begin{split}
n_{\zeta}-1&=-2\epsilon_*-\left[\frac{1}{\epsilon_*^{\phi}}\left(x_F-\frac{\sqrt{2\epsilon_*^{\phi}}}{m_P}(I_1+I_2)\right)^2 + \frac{1}{\epsilon_*^{\chi}}\left(y_F-\frac{\sqrt{2\epsilon_*^{\chi}}}{m_P}(I_3+I_4)\right)^2\right]^{-1}\\
&\times4\Bigg\{\left[1-\frac{F_*F_*''}{F_*'^2}\right]\left[4+\frac{\epsilon_*^{\chi}}{\epsilon_*^{\phi}}x_F+\frac{\epsilon_*^{\phi}}{\epsilon_*^{\chi}}y_F-\left(2+\frac{\epsilon_*^{\chi}}{\epsilon_*^{\phi}}\right)\frac{\sqrt{2\epsilon_*^{\phi}}}{m_P}(I_1+I_2)-\left(2+\frac{\epsilon_*^{\phi}}{\epsilon_*^{\chi}}\right)\frac{\sqrt{2\epsilon_*^{\chi}}}{m_P}(I_3+I_4)\right]\\
&\qquad+\left[\frac{F_*F_*''}{F_*'^2}-\frac{\eta_*^{\phi}}{2\epsilon_*^{\phi}}\right]\left[x_F^2-x_F\frac{\sqrt{2\epsilon_*^{\phi}}}{m_P}(I_1+I_2)\right]+\left[\frac{F_*F_*''}{F_*'^2}-\frac{\eta_*^{\chi}}{2\epsilon_*^{\chi}}\right]\left[y_F^2-y_F\frac{\sqrt{2\epsilon_*^{\chi}}}{m_P}(I_3+I_4)\right]\\
&\qquad-\left[J_1+J_3+\sqrt{\frac{\epsilon_*^{\chi}}{\epsilon_*^{\phi}}}(J_2+J_4)\right]\left[x_F-\frac{\sqrt{2\epsilon_*^{\phi}}}{m_P}(I_1+I_2)\right]-\left[J_6+J_8+\sqrt{\frac{\epsilon_*^{\phi}}{\epsilon_*^{\chi}}}(J_2+J_4)\right]\left[y_F-\frac{\sqrt{2\epsilon_*^{\chi}}}{m_P}(I_3+I_4)\right]\Bigg\}\, .
\end{split}
\end{align}
And finally $f_{\mathrm{NL}}^{(4)}$:
\begin{align}\label{fnl4F}
\begin{split}
\frac{6}{5}f_{\mathrm{NL}}^{(4)}&=\left[\frac{1}{\epsilon_*^{\phi}}\left(x_F-\frac{\sqrt{2\epsilon_*^{\phi}}}{m_P}(I_1+I_2)\right)^2 + \frac{1}{\epsilon_*^{\chi}}\left(y_F-\frac{\sqrt{2\epsilon_*^{\chi}}}{m_P}(I_3+I_4)\right)^2\right]^{-2}\\
&\times4\Bigg\{\left[1-\frac{F_*F_*''}{F_*'^2}\right]\Bigg[\frac{1}{\epsilon_*^{\phi}}\left(x_F-\frac{\sqrt{2\epsilon_*^{\phi}}}{m_P}(I_1+I_2)\right)^2+\frac{1}{\epsilon_*^{\chi}}\left(y_F-\frac{\sqrt{2\epsilon_*^{\chi}}}{m_P}(I_3+I_4)\right)^2\\
&\qquad\qquad\qquad\qquad\quad+\frac{2\epsilon_*}{\epsilon_*^{\phi}\epsilon_*^{\chi}}\left(x_F-\frac{\sqrt{2\epsilon_*^{\phi}}}{m_P}(I_1+I_2)\right)\left(y_F-\frac{\sqrt{2\epsilon_*^{\chi}}}{m_P}(I_3+I_4)\right)\Bigg]\\
&\quad+\left[\frac{F_*F_*''}{F_*'^2}-\frac{\eta_*^{\phi}}{2\epsilon_*^{\phi}}\right]\frac{x_F}{\epsilon_*^{\phi}}\left(x_F-\frac{\sqrt{2\epsilon_*^{\phi}}}{m_P}(I_1+I_2)\right)^2+\left[\frac{F_*F_*''}{F_*'^2}-\frac{\eta_*^{\chi}}{2\epsilon_*^{\chi}}\right]\frac{y_F}{\epsilon_*^{\chi}}\left(x_F-\frac{\sqrt{2\epsilon_*^{\chi}}}{m_P}(I_3+I_4)\right)^2\\
&\quad-\frac{1}{\epsilon_*^{\phi}}(J_1+J_3)\left(x_F-\frac{\sqrt{2\epsilon_*^{\phi}}}{m_P}(I_1+I_2)\right)^2-\frac{1}{\epsilon_*^{\chi}}(J_6+J_8)\left(y_F-\frac{\sqrt{2\epsilon_*^{\chi}}}{m_P}(I_3+I_4)\right)^2\\
&\quad-\frac{2}{\sqrt{\epsilon_*^{\phi}\epsilon_*^{\chi}}}(J_2+J_4)\left(x_F-\frac{\sqrt{2\epsilon_*^{\phi}}}{m_P}(I_1+I_2)\right)\left(y_F-\frac{\sqrt{2\epsilon_*^{\chi}}}{m_P}(I_3+I_4)\right)\\
&\quad+\frac{\epsilon_c^{\phi}\epsilon_c^{\chi}}{\epsilon_c}\frac{F_cF_*'}{F_c'F_*}\left(1-\frac{F_cF_c''}{F_c'^2}+\frac{2\eta_c^{ss}}{\epsilon_c}\right)\left[\frac{1}{\epsilon_*^{\phi}}\left(x_F-\frac{\sqrt{2\epsilon_*^{\phi}}}{m_P}(I_1+I_2)\right)-\frac{1}{\epsilon_*^{\chi}}\left(y_F-\frac{\sqrt{2\epsilon_*^{\chi}}}{m_P}(I_3+I_4)\right)\right]^2\Bigg\}\, .
\end{split}
\end{align}

For each of these quantities, if we take the limit $t_c\rightarrow t_*$ we obtain, as we should, the usual single field results:
\begin{align}
\left.P_{\zeta}\right|_{c\rightarrow *}&=\frac{W_*}{24\pi^2m_p^4\epsilon_*}\, ,\nonumber\\
\left.n_{\zeta}-1\right|_{c\rightarrow *}&=-6\epsilon_*+2\eta^{\sigma\sigma}_*\, ,\nonumber\\
\left.\frac{6}{5}f_{\mathrm{NL}}^{(4)}\right|_{c\rightarrow *}&=2\epsilon_*-\eta^{\sigma\sigma}_*\, .
\end{align}
\end{widetext}

\subsection{The Approach to Adiabaticity and the Suppression of $f_{\mathrm{NL}}^{\textrm{local}}$}
We will now apply the same analysis as in section \ref{fnlfate} to the more general potential discussed above.  This case is necessarily more complicated than both the homogeneous potential and the exponential potential due to the presence of the integrals $I_i$ and $J_i$ as well as terms such as $\frac{F_cF_*'}{F_c'F_*}$ which cannot be expressed in terms of slow roll parameters.  We will find that these details do not greatly affect the conclusions, but the argument is less straightforward in this case.

We begin by determining the approximate magnitudes of the various terms that appear in (\ref{fnl4F}) by counting factors of slow roll parameters, where as in Section \ref{fnlfate}, $\varepsilon$ denotes a general first-order slow roll parameter.  We will take the combinations $\left(x_F-\frac{\sqrt{2\epsilon_*^{\phi}}}{m_P}(I_1+I_2)\right)$ and $\left(y_F-\frac{\sqrt{2\epsilon_*^{\chi}}}{m_P}(I_3+I_4)\right)$ to be $\mathcal{O}$(1).  This is justified by the fact that if either of these combinations were much larger, say $\mathcal{O}(\varepsilon^{-1})$, then all of the terms in $f_{\mathrm{NL}}^{(4)}$ would be at most $\mathcal{O}(\varepsilon^2)$, and so $f_{\mathrm{NL}}^{(4)}$ would always be small.  On the other hand, if both of these combinations were small say $\mathcal{O}(\varepsilon)$, then the denominator of both (\ref{nsF}) and (\ref{fnl4F}) would be very small, so in this case, $f_{\mathrm{NL}}^4$ may be large, but $n_{\zeta}-1$ will also be large, and so this case is already observationally ruled out.  Given the definitions (\ref{xf}), this also constrains the combination $\frac{F_cF_*'}{F_c'F_*}$ to be $\lesssim\mathcal{O}(1)$.  With these considerations in mind, we find that the denominator of (\ref{fnl4F}) is $\mathcal{O}(\varepsilon^{-2})$ while the denominator of (\ref{nsF}) is $\mathcal{O}(\varepsilon^{-1})$.

Examining the various terms in (\ref{nsF}) we see that in order to maintain a nearly scale-invariant spectrum, we must have $\left[1-\frac{F_*F_*''}{F_*'^2}\right]$, $\left[\frac{F_*F_*''}{F_*'^2}-\frac{\eta_*^{\phi}}{2\epsilon_*^{\phi}}\right]$, $\left[\frac{F_*F_*''}{F_*'^2}-\frac{\eta_*^{\chi}}{2\epsilon_*^{\chi}}\right]$, and each of the $J_i$ $\lesssim\mathcal{O}(1)$.  The second through the sixth lines of (\ref{fnl4F}) are each $\mathcal{O}(\varepsilon^{-1})$, while the seventh line is $\mathcal{O}(\varepsilon^{-2})$.  Putting this all together, we see that the only term in $f_{\mathrm{NL}}^{(4)}$ which is not automatically slow roll suppressed is the term on the seventh line of (\ref{fnl4F}).  So if $f_{\mathrm{NL}}^{(4)}$ is going to be large, the leading contribution will be:
\begin{equation}\label{fnl4Fleading}
\frac{6}{5}f_{\mathrm{NL}}^{(4)}\sim\mathcal{O}(\varepsilon)+\mathcal{O}(1) \times \frac{\epsilon_c^{\phi}\epsilon_c^{\chi}}{\epsilon_c}\left(1-\frac{F_cF_c''}{F_c'^2}+\frac{2\eta_c^{ss}}{\epsilon_c}\right)\, .
\end{equation}

As discussed in the main text, we wish to examine the case where we pass through a phase of effectively single field inflation in order to damp away non-adiabatic fluctuations.  This requires that $\eta^{ss}$ becomes large for a finite period during inflation, while $\eta^{\sigma\sigma}$ remains small.  Recalling the definitions (\ref{slowroll}) and (\ref{etass1}) we have
\begin{align}\label{etassF}
\eta^{\sigma\sigma}=\frac{\epsilon^{\phi}\eta^{\phi}+4\frac{FF''}{F'^2}\epsilon^{\phi}\epsilon^{\chi}+\epsilon^{\chi}\eta^{\chi}}{\epsilon}\, ,\nonumber\\
\eta^{ss}=\frac{\epsilon^{\chi}\eta^{\phi}-4\frac{FF''}{F'^2}\epsilon^{\phi}\epsilon^{\chi}+\epsilon^{\phi}\eta^{\chi}}{\epsilon}\, .
\end{align}
The term $\frac{4}{\epsilon}\frac{FF''}{F'^2}\epsilon^{\phi}\epsilon^{\chi}$ cannot give the dominant contribution to $\eta^{ss}$ when $\eta^{ss}$ becomes large, because the same term appears in $\eta^{\sigma\sigma}$ which must remain small otherwise inflation will quickly end.  Here we have assumed that the potential is not so finely-tuned as to create a cancellation which would keep $\eta^{\sigma\sigma}$ small while $\frac{FF''}{F'^2}$ is large.  As a result, we must have either $\eta^{\phi}\gtrsim\frac{\epsilon}{\epsilon^{\chi}}$ or $\eta^{\chi}\gtrsim\frac{\epsilon}{\epsilon^{\phi}}$ in order to damp away non-adiabatic fluctuations, while $\frac{4}{\epsilon}\frac{FF''}{F'^2}\epsilon^{\phi}\epsilon^{\chi}\sim\mathcal{O}(\varepsilon)$.  This means that only the term proportional to $\eta^{ss}$ will be important if $f_{\mathrm{NL}}^{(4)}$ is large.

From this point forward, the argument is precisely the same as is given in section \ref{fnlfate}, and we find that for $\eta^{ss}>1$
\begin{align}\label{dampingF}
|\delta s|&\sim \textrm{Exp}\left[-\frac{3}{2}\int H\textrm{d}t\right] \\
f_{\mathrm{NL}}^{(4)}&\sim\mathcal{O}(\varepsilon_*)+\mathcal{O}(1)\times\eta^{ss}\textrm{Exp}\left[-2\int C_{\eta}H\eta^{ss}\textrm{d}t\right]\, .
\end{align}
$C_{\eta}$ is a number which is always greater than 1, and depends on the direction of the effective inflaton during this phase (as in section \ref{fnlfate}). Therefore, for any potential of the form $W(\phi,\chi)=F(U(\phi)+V(\chi))$, we find that $f_{\mathrm{NL}}^{(4)}$ will always be small upon entering the purely adiabatic solution after passing through a phase of effectively single field inflation.

\bibliography{nongauss}

\end{document}